\documentclass[10pt,twocolumn,twoside]{IEEEtran}

\usepackage{epsfig,algorithm,url,subfigure}
\usepackage{amsmath,amssymb}
\title{
Particle Filtering for Large Dimensional State Spaces with Multimodal Observation Likelihoods
}
\author{Namrata Vaswani
\thanks{N. Vaswani is with the Dept. of Electrical and Computer Engineering, Iowa State University, Ames, IA 50011 (email: namrata@iastate.edu). }
\thanks{A part of this paper is contained in \cite{icassp07}. Some initial ideas appeared in \cite{pap1}.}
}

\begin{document}
\maketitle
\date{}

\newcommand{\Dnum}{D_{num}}
\newcommand{\pss}{p^{**,i}}
\newcommand{\fr}{f_{r}^i}

\newcommand{\A}{{\cal A}}
\newcommand{\Z}{{\cal Z}}
\newcommand{\B}{{\cal B}}
\newcommand{\R}{{\cal R}}
\newcommand{\reg}{{\cal G}}
\newcommand{\const}{\mbox{const}}

\newcommand{\trace}{\mbox{trace}}

\newcommand{\hsim}{{\hspace{0.0cm} \sim  \hspace{0.0cm}}}
\newcommand{\he}{{\hspace{0.0cm} =  \hspace{0.0cm}}}

\newcommand{\vect}[2]{\left[\begin{array}{cccccc}
     #1 \\
     #2
   \end{array}
  \right]
  }

\newcommand{\matr}[2]{ \left[\begin{array}{cc}
     #1 \\
     #2
   \end{array}
  \right]
  }
\newcommand{\vc}[2]{\left[\begin{array}{c}
     #1 \\
     #2
   \end{array}
  \right]
  }

\newcommand{\gdot}{\dot{g}}
\newcommand{\Cdot}{\dot{C}}
\newcommand{\re}{\mathbb{R}}
\newcommand{\n}{{\cal N}}  
\newcommand{\N}{{\overrightarrow{\bf N}}}  
\newcommand{\chat}{\tilde{C}_t}
\newcommand{\chati}{\chat^i}

\newcommand{\cmin}{C^*_{min}}
\newcommand{\twi}{\tilde{w}_t^{(i)}}
\newcommand{\twj}{\tilde{w}_t^{(j)}}
\newcommand{\wi}{{w}_t^{(i)}}
\newcommand{\twio}{\tilde{w}_{t-1}^{(i)}}

\newcommand{\tWi}{\tilde{W}_n^{(m)}}
\newcommand{\tWj}{\tilde{W}_n^{(k)}}
\newcommand{\Wi}{{W}_n^{(m)}}
\newcommand{\tWio}{\tilde{W}_{n-1}^{(m)}}

\newcommand{\ds}{\displaystyle}

\newcommand{\SAR}{S$\!$A$\!$R }
\newcommand{\MAR}{MAR}
\newcommand{\MMRF}{MMRF}
\newcommand{\AR}{A$\!$R }
\newcommand{\GMRF}{G$\!$M$\!$R$\!$F }
\newcommand{\DTM}{D$\!$T$\!$M }
\newcommand{\MSE}{M$\!$S$\!$E }
\newcommand{\RCS}{R$\!$C$\!$S }
\newcommand{\uomega}{\underline{\omega}}
\newcommand{\y}{v}
\newcommand{\x}{w}
\newcommand{\lu}{\mu}
\newcommand{\g}{g}
\newcommand{\s}{{\bf s}}
\newcommand{\bft}{{\bf t}}
\newcommand{\refmap}{{\cal R}}
\newcommand{\totrefl}{{\cal E}}
\newcommand{\beq}{\begin{equation}}
\newcommand{\eeq}{\end{equation}}
\newcommand{\bdm}{\begin{displaymath}}
\newcommand{\edm}{\end{displaymath}}
\newcommand{\hatz}{\hat{z}}
\newcommand{\hatu}{\hat{u}}
\newcommand{\tilz}{\tilde{z}}
\newcommand{\tilu}{\tilde{u}}
\newcommand{\hhatz}{\hat{\hat{z}}}
\newcommand{\hhatu}{\hat{\hat{u}}}
\newcommand{\tilc}{\tilde{C}}
\newcommand{\hatc}{\hat{C}}
\newcommand{\tim}{n}

\newcommand{\ssp}{\renewcommand{\baselinestretch}{1.0}}
\newcommand{\defd}{\mbox{$\stackrel{\mbox{$\triangle$}}{=}$}}
\newcommand{\goes}{\rightarrow}
\newcommand{\tends}{\rightarrow}
\newcommand{\defn}{\triangleq} 
\newcommand{\se}{&=&}
\newcommand{\sdefn}{& \defn  &}
\newcommand{\sle}{& \le &}
\newcommand{\sge}{& \ge &}
\newcommand{\plusminus}{\stackrel{+}{-}}
\newcommand{\Ey}{E_{Y_{1:t}}}
\newcommand{\ey}{E_{Y_{1:t}}}

\newcommand{\equivto}{\mbox{~~~which is equivalent to~~~}}
\newcommand{\nonzero}{i:\pi^n(x^{(i)})>0}
\newcommand{\nonzeroc}{i:c(x^{(i)})>0}

\newcommand{\supn}{\sup_{\phi:||\phi||_\infty \le 1}}
\newtheorem{theorem}{Theorem}
\newtheorem{lemma}{Lemma}
\newtheorem{corollary}{Corollary}
\newtheorem{definition}{Definition}
\newtheorem{remark}{Remark}
\newtheorem{example}{Example}
\newtheorem{ass}{Assumption}
\newtheorem{fact}{Fact}
\newtheorem{heuristic}{Heuristic}
\newcommand{\eps}{\epsilon}
\newcommand{\bd}{\begin{definition}}
\newcommand{\ed}{\end{definition}}
\newcommand{\udq}{\underline{D_Q}}
\newcommand{\td}{\tilde{D}}
\newcommand{\epsinv}{\epsilon_{inv}}
\newcommand{\al}{\mathcal{A}}

\newcommand{\bfx} {\bf X}
\newcommand{\bfy} {\bf Y}
\newcommand{\bfz} {\bf Z}
\newcommand{\ddas}{\mbox{${d_1}^2({\bf X})$}}
\newcommand{\ddbs}{\mbox{${d_2}^2({\bfx})$}}
\newcommand{\dda}{\mbox{$d_1(\bfx)$}}
\newcommand{\ddb}{\mbox{$d_2(\bfx)$}}
\newcommand{\xinc}{{\bfx} \in \mbox{$C_1$}}
\newcommand{\eqa}{\stackrel{(a)}{=}}
\newcommand{\eqb}{\stackrel{(b)}{=}}
\newcommand{\eqe}{\stackrel{(e)}{=}}
\newcommand{\leqc}{\stackrel{(c)}{\le}}
\newcommand{\leqd}{\stackrel{(d)}{\le}}

\newcommand{\leqa}{\stackrel{(a)}{\le}}
\newcommand{\leqb}{\stackrel{(b)}{\le}}
\newcommand{\leqe}{\stackrel{(e)}{\le}}
\newcommand{\leqf}{\stackrel{(f)}{\le}}
\newcommand{\leqg}{\stackrel{(g)}{\le}}
\newcommand{\leqh}{\stackrel{(h)}{\le}}
\newcommand{\leqi}{\stackrel{(i)}{\le}}
\newcommand{\leqj}{\stackrel{(j)}{\le}}

\newcommand{\w}{{W^{LDA}}}
\newcommand{\halpha}{\hat{\alpha}}
\newcommand{\hsigma}{\hat{\sigma}}
\newcommand{\slmax}{\sqrt{\lambda_{max}}}
\newcommand{\slmin}{\sqrt{\lambda_{min}}}
\newcommand{\lmax}{\lambda_{max}}
\newcommand{\lmin}{\lambda_{min}}

\newcommand{\da} {\frac{\alpha}{\sigma}}
\newcommand{\chka} {\frac{\check{\alpha}}{\check{\sigma}}}
\newcommand{\sumo}{\sum _{\underline{\omega} \in \Omega}}
\newcommand{\distance}{d\{(\hatz _x, \hatz _y),(\tilz _x, \tilz _y)\}}
\newcommand{\col}{{\rm col}}
\newcommand{\rcs}{\sigma_0}
\newcommand{\CalR}{{\cal R}}
\newcommand{\df}{{\delta p}}
\newcommand{\dq}{{\delta q}}
\newcommand{\dZ}{{\delta Z}}
\newcommand{\pprime}{{\prime\prime}}

\newcommand{\vn}{N}

\newcommand{\bv}{\begin{vugraph}}
\newcommand{\ev}{\end{vugraph}}
\newcommand{\bi}{\begin{itemize}}
\newcommand{\ei}{\end{itemize}}
\newcommand{\ben}{\begin{enumerate}}
\newcommand{\een}{\end{enumerate}}
\newcommand{\be}{\protect\[}
\newcommand{\ee}{\protect\]}
\newcommand{\bean}{\begin{eqnarray*} }
\newcommand{\eean}{\end{eqnarray*} }
\newcommand{\bea}{\begin{eqnarray} }
\newcommand{\eea}{\end{eqnarray} }
\newcommand{\nn}{\nonumber}
\newcommand{\ba}{\begin{array} }
\newcommand{\ea}{\end{array} }
\newcommand{\ep}{\mbox{\boldmath $\epsilon$}}
\newcommand{\epp}{\mbox{\boldmath $\epsilon '$}}
\newcommand{\Lep}{\mbox{\LARGE $\epsilon_2$}}
\newcommand{\und}{\underline}
\newcommand{\pdif}[2]{\frac{\partial #1}{\partial #2}}
\newcommand{\odif}[2]{\frac{d #1}{d #2}}
\newcommand{\dt}[1]{\pdif{#1}{t}}
\newcommand{\urho}{\underline{\rho}}

\newcommand{\spc}{{\cal S}}
\newcommand{\tspc}{{\cal TS}}

\newcommand{\uv}{\underline{v}}
\newcommand{\us}{\underline{s}}
\newcommand{\uc}{\underline{c}}
\newcommand{\utheta}{\underline{\theta}^*}
\newcommand{\ualpha}{\underline{\alpha^*}}

\newcommand{\uxy}{\underline{x}^*}
\newcommand{\uxyj}{[x^{*}_j,y^{*}_j]}
\newcommand{\arcl}[1]{arclen(#1)}
\newcommand{\one}{{\mathbf{1}}}

\newcommand{\uxyjt}{\uxy_{j,t}}
\newcommand{\E}{\mathbb{E}}

\newcommand{\rhomat}{\left[\begin{array}{c}
                        \rho_3 \ \rho_4 \\
                        \rho_5 \ \rho_6
                        \end{array}
                   \right]}
\newcommand{\deltat}{\tau} 
\newcommand{\deltatt}{\Delta t_1}
\newcommand{\ceil}[1]{\ulcorner #1 \urcorner}

\newcommand{\xxi}{x^{(i)}}
\newcommand{\txi}{\tilde{x}^{(i)}}
\newcommand{\txj}{\tilde{x}^{(j)}}

\newcommand{\mi}[1]{{#1}^{(m,i)}}

\newcommand{\Section}[1]{\section{#1}}

\newcommand{\Subsection}[1]{\subsection{#1}}

\begin{figure*}[t!]
\centerline{
\subfigure[Bimodal OL, Narrow STP: Unimodal $p^*$] 
{
\label{works}
\vspace{-0.3in}
\epsfig{file = 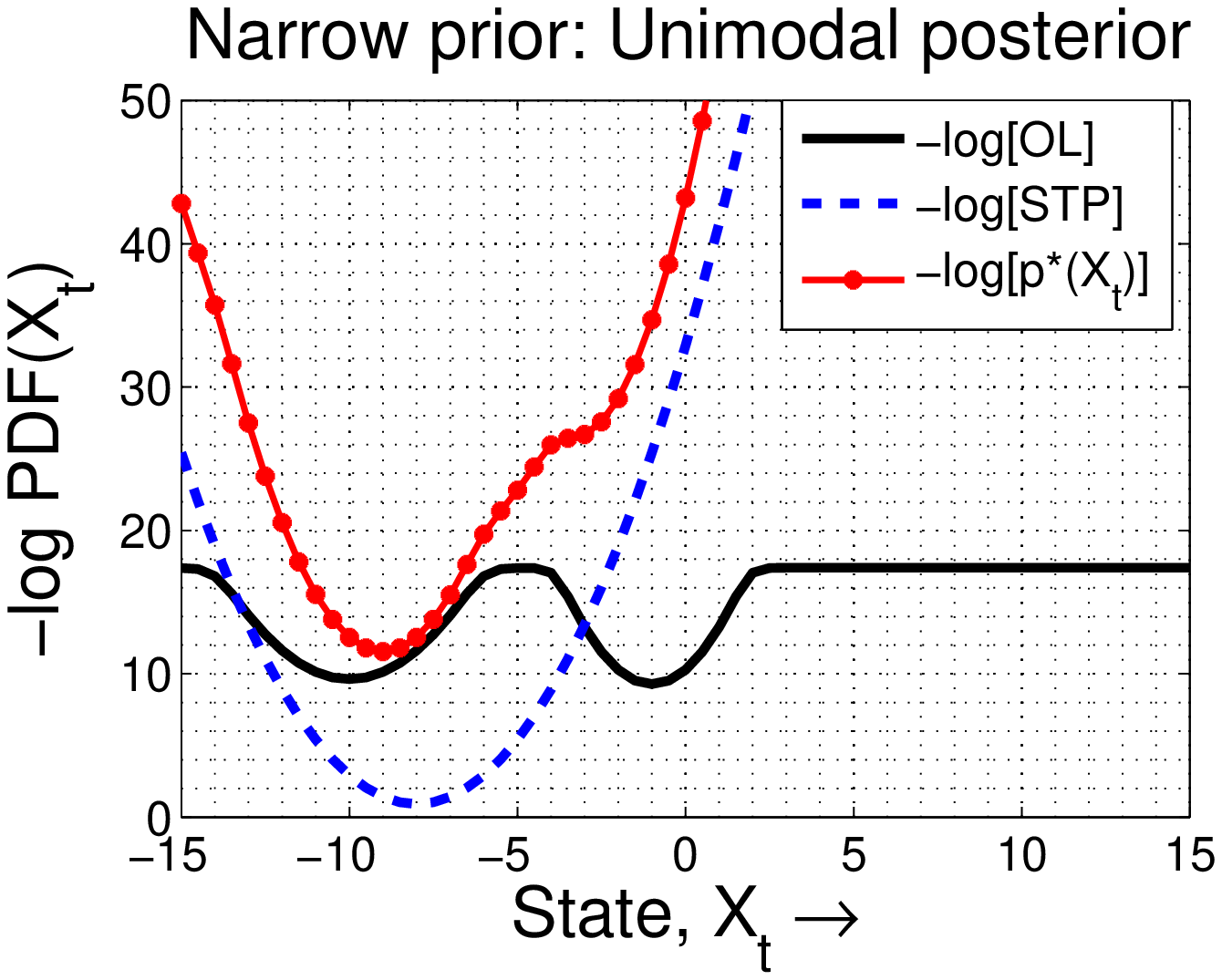, width=5.5cm,height=3.0cm}
}
\subfigure[Bimodal OL, Broad STP: Bimodal $p^*$]  
{
\label{fails}
\vspace{-0.3in}
\epsfig{file = 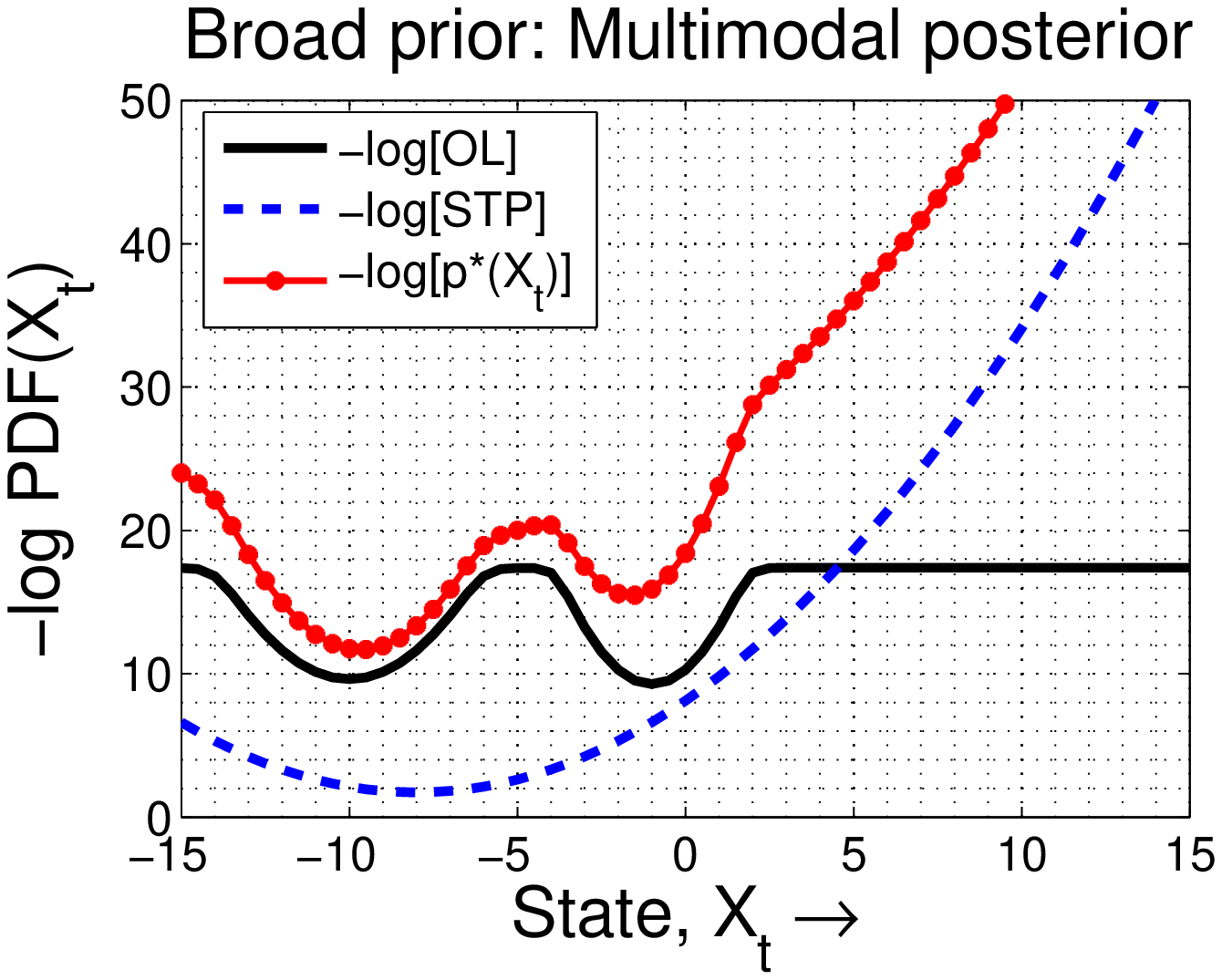, width=5.5cm,height=3.0cm}
}
\subfigure[Heavy-tailed OL, Broad STP: Bimodal $p^*$] 
{
\label{fails2}
\vspace{-0.3in}
\epsfig{file = 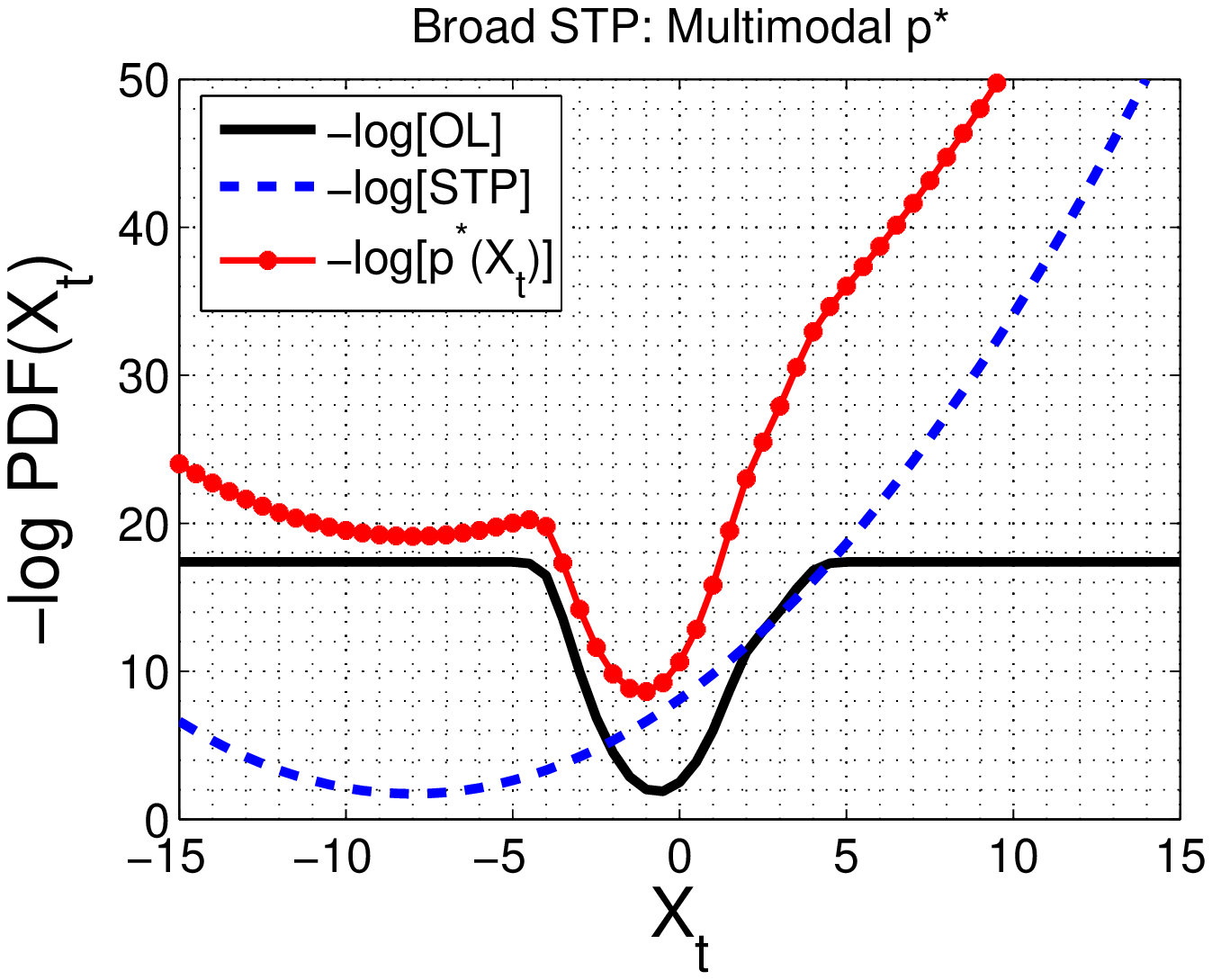, width=5.5cm,height=3.0cm}
}
}
\caption{{\small
Demonstrating the effect of multimodal or heavy-tailed OL and broad STP for a $M=1$ dimensional version of Example \ref{temptrack} with temperature independent failure. $X_t$ is temperature.
The STP is $\n(X_{t-1},\sigma_{sys}^2)$, i.e. Example \ref{temptrack} with $a=0$.
Fig. \ref{works}: One out of $J=2$ sensors fails (bimodal OL) but narrow enough STP ($\sigma_{sys}^2=1$). So $p^*$ is unimodal.
Fig. \ref{fails}: One out of $J=2$ sensors fails (bimodal OL) and broad STP ($\sigma_{sys}^2=5$). So $p^*$ is bimodal.
Fig. \ref{fails2}: Estimating temperature but with $J=1$ sensor and broad STP ($\sigma_{sys}^2=5$). When the sensor fails, the OL is heavy-tailed and peaks at the wrong mode. Thus $p^*$ is bimodal with the wrong mode being the strong one. Note that the correct mode is so weak it may get missed in numerical computations.
}}
\label{kfpmtfail}
\end{figure*}

\begin{abstract} 
We study efficient importance sampling techniques for particle filtering (PF) when either (a) the observation likelihood (OL) is frequently multimodal or heavy-tailed, or (b) the state space dimension is large or both. When the OL is multimodal, but the state transition pdf (STP) is narrow enough, the optimal importance density  is usually unimodal. Under this assumption, many techniques have been proposed. But when the STP is broad, this assumption does not hold. We study how existing techniques can be generalized to situations where the optimal importance density is multimodal, but is unimodal conditioned on a part of the state vector.

Sufficient conditions to test for the unimodality of this conditional posterior are derived. Our result is directly extendable to testing for unimodality of any posterior.%

The number of particles, N, to accurately track using a PF increases with state space dimension, thus making any regular PF impractical for large dimensional tracking problems. But in most such problems, most of the state change occurs in only a few dimensions, while the change in the rest of the dimensions is small. Using this property, we propose to replace importance sampling from a large part of the state space (whose conditional posterior is narrow enough) by just tracking the mode of the conditional posterior. This introduces some extra error, but it also greatly reduces the importance sampling dimension. The net effect is much smaller error for a given N, especially when the available N is small. 
An important class of large dimensional problems with multimodal OL is tracking spatially varying physical quantities such as temperature or pressure in a large area using a network of sensors which may be nonlinear and/or may have non-negligible failure probabilities. Improved performance of our proposed algorithms over existing PFs is demonstrated for this problem.
\end{abstract}


\section{Introduction}
Tracking is the problem of causally estimating a hidden state sequence, $\{X_t\}$, from a sequence of noisy and possibly nonlinear observations, $\{Y_t\}$ that satisfy the Hidden Markov Model (HMM) assumption.
A tracker recursively computes (or approximates) the ``posterior" at time $t$, using the posterior at $t-1$ and the current observation $Y_t$. For nonlinear and/or non-Gaussian state space models, the posterior cannot be computed exactly. But, it can be efficiently approximated using a sequential Monte Carlo method called particle filtering (PF) \cite{gordon,tutorial,mcip}. A PF outputs at each time $t$, a cloud of $N$ weighted particles whose empirical measure closely approximates the true posterior for large $N$. A generic PF is summarized in Algorithm \ref{pf}.
%
%
%
 There are two main issues in PF design: (a) choice of importance sampling density that reduces the variance of the particle weights and thus improves ``effective particle size" \cite{doucet} and (b) choice of  resampling techniques that improve effective particle size while not significantly increasing ``particle impoverishment" \cite{tutorial}.
Some solutions for (b) are  \cite[Ch.~13]{mcip},\cite{cormick,auxPF}. 
Our focus is on designing efficient importance densities and analyzing the assumptions under which they work, when either or both of the following occur:
\ben
\item The observation likelihood (OL) is frequently multimodal or heavy-tailed (or most generally, not strongly log-concave) as a function of the state and the state transition prior (STP) is broad. 

\item State space dimension is large (typically more than 10 or 12). It is well known \cite{gordon,gustaffson} that the number of particles for a given tracking accuracy increases with state space dimension. This makes any regular PF impractical for large dimensional state spaces (LDSS).%

\een

\bd[Multimodal (or heavy-tailed) OL] 
 refers to the OL, $p(Y_t|X_t)$, having multiple local maxima (or a heavy tail) as a function of the state, $X_t$, for a given observation, $Y_t$. 
An example is the observation model for the nonstationary growth model of \cite{gordon}: $Y_t = X_t^2 + w_t$. Here, the OL is bimodal with modes at $X_t = \pm \sqrt{Y_t}$ whenever $Y_t$ is significantly positive. Another example is the clutter model of \cite{condensation}.%
%
\ed

Other examples are as follows. Consider tracking spatially varying temperature change using a network of sensors (see Example \ref{temptrack}). Whenever one or more sensors fail (e.g. due to a large unmodeled disturbance or some other damage), the OL is often heavy-tailed or multimodal (see Fig. \ref{kfpmtfail}). The models of  Example \ref{temptrack} are also similar to the commonly used clutter model in radar based target tracking applications or in contour tracking applications, e.g. Condensation \cite{condensation}, and to outlier noise models used in other visual tracking problems \cite{cdc06journal} or in aircraft navigation problems \cite{gustaffson}.
Another reason for OL multimodality is having a sensor that measures a nonlinear (many-to-one) function of the actual temperature. For e.g., the growth model of \cite{gordon}. Another many-to-one example is when the observation is a product of functions of two subsets of states plus noise, for e.g. bearings-only tracking \cite{gordon} or illumination and motion tracking \cite{kalecvpr,kaleicassp}.

Note that even though our work was motivated by tracking problems with frequently multimodal OL, it is equally well  {\em applicable to any problem where the posterior is often multimodal} (e.g. due to nonlinearities in the system model), {\em but is unimodal conditioned on a part of the state space}.%

Large dimensional state spaces (LDSS) occur in tracking time-varying random fields, such as temperature or pressure, at a large number of nodes using a network of sensors \cite{mouraRC, moura_network} (applications in environment monitoring and weather forecasting); in tracking AR parameters for noisy speech \cite{parcor}; and in visual tracking problems such as tracking deforming contours \cite{wilsky,YezziSoatto2004,keriven,cdc06journal}, tracking spatially varying illumination change \cite{kalecvpr,kaleicassp} or tracking sets of ``landmark" points \cite{asilomar}.
In all of the above problems, {\em at any time, ``most state change" occurs in a small number of dimensions, while the change in the rest of the state space is small.} We call this the {\em ``LDSS property}.
The LDSS property is related to, but different from, the assumption used by dimension reduction techniques such as Principal Components Analysis (PCA). If $X_t$ is a stationary large dimensional time series, or if $X_t$ projected along a large part of the state space is asymptotically stationary,  PCA can be used for dimension reduction. Under a similar assumption, another PF has been recently proposed \cite{chorin}. But if $X_t$ follows a random walk model (the increments, $X_t-X_{t-1}$, are stationary) in all dimensions, one cannot simply eliminate the low variance directions of $X_t-X_{t-1}$, or use \cite{chorin}. This is because the variance of $X_t$ even along these directions will be significant as $t$ increases.

A generic PF is summarized in Algorithm \ref{pf}. The most commonly used importance sampling density is the STP \cite{gordon}. This assumes nothing and is easiest to implement. But since this does not use knowledge of the observation, the weights' variance can be large (particularly when the STP is broad compared to the OL), resulting in lower effective particle sizes \cite{tutorial}. The ``optimal" importance density \cite{doucet}, i.e. one that minimizes the  variance of weights conditioned on past particles and observations until $t$, is the posterior conditioned on the previous state, denoted $p^*$. When $p^*$ is unimodal (at least approximately), PF-Doucet \cite{doucet} approximates it by a Gaussian about its mode (Laplace's approximation) and importance samples from the Gaussian.  Laplace's approximation has also been used for approximating posteriors in different contexts earlier \cite{mosteller64,lindley80,tierney86}. Other work in PF literature that also implicitly assumes that $p^*$ is unimodal includes \cite{tutorial,upf,volkan_is}. 
%
When the OL is multimodal, $p^*$ will be unimodal only if the STP is unimodal and narrow enough (see Fig. \ref{kfpmtfail}). In many situations, especially for LDSS problems, this does not hold. We develop the PF with Efficient IS (PF-EIS) algorithm to address such situations. PF-EIS assumes unimodality of $p^*$ conditioned on a few states which we call {\em ``multimodal states"}.

 Sufficient conditions to test for the unimodality of this conditional posterior are derived in Theorem \ref{unimodthm}. To the best of our knowledge, such a result has not been proved earlier. It is equally applicable to test for unimodality of any posterior.

When in addition to multimodality, the state space space dimension is also large (typically more than 10 or 12), the number of particles required for reasonable accuracy is very large \cite{gordon,gustaffson} and this makes a regular PF impractical. One solution that partially addresses this issue is \cite[Ch~13]{mcip} or \cite{cormick} which propose to  resample more than once within a time interval. But more resampling results in more particle impoverishment \cite{tutorial}. When the state space model is conditionally linear-Gaussian, or when many states can be vector quantized into a few discrete centers (need to know the centers a-priori), Rao Blackwellization (RB-PF) \cite{chen_liu,gustaffson} can be used. In general, neither assumption may hold. But when the LDSS property holds, it is possible to split the state space in such a way that the conditional posterior of a part of it is quite narrow, besides being unimodal. If it is narrow enough, importance sampling (IS) from this part of the state space can be replaced by just tracking the mode of the conditional posterior (mode tracking (MT)). The resulting algorithm is called PF-EIS-MT. MT introduces some extra error. But it greatly reduces the IS dimension. The net effect, is 
 that a much smaller number of particles are required to achieve a given error, thus making PF practical for LDSS problems.


In summary, {\em our contributions are} (a) two efficient algorithms for multimodal and large dimensional problems, (PF-EIS and PF-EIS-MT); and (b) a set of sufficient conditions to test for unimodality of the conditional posterior (Theorem \ref{unimodthm}) and  heuristics based on it to split the state space in the most efficient way.  PF-EIS and Theorem \ref{unimodthm} are derived in Sec. \ref{pfeissec}.  A generic LDSS model is introduced in Sec. \ref{ldssmodel}. Practical ways of choosing the ``multimodal states" are discussed in Sec. \ref{effbasis}. PF-EIS-MT and PF-MT are introduced in Sec. \ref{pfmtsec}. Relation to existing work is described in Sec. \ref{relwork}.
In Sec. \ref{sims}, we given extensive simulation results comparing our methods with existing work for the temperature field tracking problem.
Conclusions and open issues are presented in Sec. \ref{discussion}.

\begin{algorithm*}[t] 
\caption{{\small \em Generic PF. Going from $\pi_{t-1}^N$ to $\pi_t^N(X_t) \defn \sum_{i=1}^N \wi \delta(X_{t} - X_{t}^{i})$ (note $\delta(X-a)$ denotes a Dirac delta function at $a$)}}
\label{pf}
A PF starts with sampling $N$ times from $\pi_0$ at $t=0$ to approximate it by $\pi_t^N(X_0)$. For each $t>0$, it approximates the Bayes recursion for going from $\pi_{t-1}^N$ to $\pi_t^N$ using sequential importance sampling. This consists of the following 3 steps:
\ben
\item
 {\em Importance Sample (IS):} For $i=1,2...N$, Sample $X_t^{i} \sim q(X_t^{i})$. The IS density, $q$, can depend on $X_{1:t-1}^{i}$, $Y_{1:t}$.%
\label{isstep}
\item
 {\em Weight:} For $i=1,2...N$, compute the  weights:  $w_t^{i} = \frac{\tilde{w}_t^{i}}{\sum_{j=1}^N \tilde{w}_t^{(j)} }$, where $\tilde{w}_t^{i} = w_{t-1}^{i} \frac{p(Y_t|X_t^{i}) p(X_t^{i}|X_{t-1}^{i})}{q(X_t^{i})}$.
\item
{\em Resample:}   Replicate particles in proportion to their weights \& reset $w_t^{i}$ for all $i$ \cite{tutorial}. Set $t \leftarrow t+1$ \& go to step \ref{isstep}.
\een
\end{algorithm*} 

%
%
%

\setlength{\arraycolsep}{0.05cm}

\section{PF-EIS: PF-Efficient Importance Sampling}
\label{pfeissec}
We  denote the probability density function (pdf) of a random vector $\mathbf{X}$, $f_{\mathbf{X}}(X)$, using the notation $p(X)$ and we denote the conditional pdf, $f_{\mathbf{X|Y}}(X|Y)$, by $p(X|Y)$.
Consider tracking a hidden sequence of states $X_t$ from a sequence of observations $Y_t$ which satisfy the HMM property:%
\begin{ass}[HMM]
For each $t$,
\ben
\item  The dependence $X_{t-1} \rightarrow X_t$ is Markovian, with state transition pdf (STP), $p(X_t|X_{t-1})$.
\item Conditioned on $X_t$, $Y_t$ is independent of past and future states and observations. The observation likelihood (OL) is $p(Y_t|X_t)$.%
\een
\label{hmmass}
\end{ass}
A generic particle filter (PF) is summarized in Algorithm \ref{pf}.%

\subsection{PF-EIS Algorithm}
Consider designing a PF for a given state space model. 
The optimal importance sampling density \cite{doucet} is $p(X_t|X_{t-1}^{i},Y_t) \defn p^*(X_t)$. In most cases, this cannot be computed analytically~\cite{doucet}.  {\em If $p^*$ is unimodal (at least approximately)}, \cite{doucet} suggests approximating it by a Gaussian about its mode and sampling from it (Laplace's approximation \cite{tierney86}).
But, when the OL is multimodal, or heavy-tailed, or otherwise not strongly log-concave, $p^*$ will be unimodal only if the STP is unimodal and narrow enough and the predicted state particle is near enough to an OL mode (see Fig. \ref{kfpmtfail}). In many situations, this may not hold in all dimensions.
But in most such situations, the STP is broad and/or multimodal in only a few directions of the state space which we call the {\em ``multimodal" directions}. It can be shown that if the STP is unimodal and narrow enough in the rest of the directions, $p^*$ will be unimodal conditioned on the {\em ``multimodal states"} (Theorem \ref{unimodthm}).
When this holds, we propose to split the state vector as $X_t = [X_{t,s} ; X_{t,r}]$ in such a way that $X_{t,s}$ contains the minimum number of dimensions for which $p^*$ is unimodal conditioned on it, i.e.
\bea
p^{**,i}(X_{t,r}) \defn p^*(X_t|X_{t,s}^i) = p(X_{t,r}|X_{t-1}^{i},X_{t,s}^i, Y_t)
\label{defpss}
\eea
is unimodal.
We sample $X_{t,s}$ from its STP (to sample the possibly multiple modes of $p^*$), and use Laplace's approximation to approximate $p^{**,i}$ and sample $X_{t,r}$ from it, i.e. sample $X_{t,r}^i$ from $\n(m_t^i, \Sigma_{IS}^i)$ where
\bea
m_t^i \se m_t^i(X_{t-1}^i,X_{t,s}^i,Y_t) \defn \min_{X_{t,r}}  L^i(X_{t,r}), \ \text{where,}  \nn \\
\Sigma_{IS}^i \sdefn [(\nabla^2 L^i)(m_t^i)]^{-1} \ \ \nn \\
L^i(X_{t,r}) \sdefn -\log[ p^{**,i}(X_{t,r})] + \const
\label{defsigis}
\eea
$\nabla^2 L^i$ denotes the Hessian of $L^i$. The weighting step also changes to satisfy the principle of importance sampling.
The complete algorithm is given in Algorithm \ref{pfeis}. We call it PF with Efficient Importance Sampling (PF-EIS).
As we shall see later, it is very expensive to exactly verify the unimodality conditions of Theorem \ref{unimodthm}. But even if $X_{t,s}$ is chosen so that $\pss$ is unimodal for most particles and at most times (i.e. is unimodal with high probability), the proposed algorithm works well. This can be seen from the simulation results of Sec. \ref{sims}.

\begin{algorithm*}[t!]   
\caption{{\small \bf PF-EIS. Going from $\pi_{t-1}^N$ to $\pi_t^N(X_t) = \sum_{i=1}^N \wi \delta(X_{t} - X_{t}^{i})$, \ $X_{t}^{i} = [X_{t,s}^{i},X_{t,r}^{i}]$  }}
\label{pfeis}
\ben
\item
{\em Importance Sample $X_{t,s}$: } $\forall i$, sample $X_{t,s}^{i} \sim p(X_{t,s}^i|X_{t-1}^i)$.
\label{isxts}

\item {\em Efficient Importance Sample $X_{t,r}$: } $\forall i$, sample $X_{t,r}^i \sim \n(X_{t,r}^i; m_t^{i}, \ \Sigma_{IS}^{i})$. Here $m_t^i(X_{t-1}^i,X_{t,s}^i,Y_t) =  \arg \min_{X_{t,r}} L^i(X_{t,r})$ and $\Sigma_{IS}^{i}\defn (\nabla^2 L^i(m_t^{i}))^{-1}$ and $L^i$ is defined in (\ref{defL}).
\label{isxtr}

\item {\em Weight: }  $\forall i$, compute $w_t^i = \frac{\tilde{w}_t^{i} }{ \sum_{j=1}^N \tilde{w}_t^{j}}$ where  $\tilde{w}_t^{i} = w_{t-1}^{i} \frac{p(Y_t|X_t^{i}) p(X_{t,r}^{i}|X_{t-1}^i, X_{t,s}^i)}{\n(X_{t,r}^{i}; \ m_t^{i}, \ \Sigma_{IS}^{i})}$  where  $X_t^i=[X_{t,s}^i,X_{t,r}^i]$.
\item {\em Resample \cite{tutorial}.}  Set $t \leftarrow t+1$ \& go to step \ref{isxts}.
\label{weight}
\een
\end{algorithm*}

\subsection{Conditional Posterior Unimodality} 
\label{unimodanal}
We derive sufficient conditions for unimodality of the conditional posterior, $\pss$. 
 Let $\dim(X_{t,s}) \defn K$, $\dim(X_{t,r}) \defn M_r$, $\dim(X_t) \defn M = K + M_r$. Because of the HMM structure,
\bea
p^{**,i}(X_{t,r}) 
\se  \zeta p(Y_t|X_{t,s}^i, X_{t,r}) p(X_{t,r}|X_{t-1}^{i},X_{t,s}^i)
\label{pssprop}
\eea
where $\zeta$ is a proportionality constant.
\setlength{\arraycolsep}{0.05cm}
\begin{definition}
We first define a few terms and symbols.
\ben
\item The notation $A>0$ ($A \ge 0$) where $A$ is a square matrix means that $A$ is {\em positive definite (positive semi-definite)}. Also, $A>B$ ($A \ge B$) means $A-B > 0$ ($A-B \ge 0$).

\item The term {\em ``minimizer"} refers to the unconstrained local minimizer of a function, i.e. a point $x_0$ s.t. $f(x_0) \le f(x)$ $\forall \ x$ in its neighborhood. Similarly for ``maximizer".

\item A twice differentiable function, $f(x)$, is {\em strongly convex} in a region $\R$, if there exists an $m>0$ s.t. at all points, $x \in \R$, the Hessian $\nabla^2 f(x) \ge m I$. 
If $f$ is strongly convex in $\R$, it has at most one minimizer in $\R$ and it lies in the interior of $\R$. If $f$ is strongly-convex on $\re^M$, then it has exactly one (finite) minimizer. 

\item A function is {\em strongly log-concave} if its negative log is strongly convex. An example is a Gaussian pdf.

\item Since a pdf is an integrable function, it will always have at least one (finite) maximizer. Thus a pdf having at most one maximizer is equivalent to it being {\em unimodal}.

\item The symbol $\E[.]$ denotes expected value. 
\item We denote the $-\log$ of OL using the symbol $E_{Y_t}$, i.e.
\bea
E_{Y_t}(X_t) \defn -\log p(Y_t|X_t) + \const
\label{defE}
\eea
\item We denote the $-\log$ of the STP of $X_{t,r}$ as
\bea
D^i(X_{t,r}) \sdefn -\log p(X_{t,r}|X_{t-1}^i,X_{t,s}^i) + \const
\label{defD}
\eea
\item When the STP of $X_{t,r}$ is strongly log-concave (assumed in Theorem \ref{unimodthm}), we denote its unique mode by
\bea
f_r^i \sdefn f_r(X_{t-1}^i,X_{t,s}^i) = \arg \max_{X_{t,r}} p(X_{t,r}|X_{t-1}^i,X_{t,s}^i) \ \ \ \ \
\label{deffr}
\eea

\item $[z]_p$ or $z_p$ denotes the $p^{th}$ coordinate of a vector, $z$.

\item $\max_p$ is often used in place of $\max_{p=1,2,\dots M_r}$.
\een
\end{definition}
\vspace{0.1in}
Combining (\ref{pssprop}), (\ref{defE}) and (\ref{defD}), $L^i(X_{t,r})$ can be written as
\bea
\label{defL}
L^i(X_{t,r}) \se E_{Y_t}(X_{t,s}^i,X_{t,r}) +  D^i(X_{t,r})
\eea
 Now, $\pss(X_{t,r})$ will be unimodal if and only if we can show that $L^i$ has at most one minimizer. We derive a set of sufficient conditions on $E_{Y_t}$, $D^i$ and $\fr$ to ensure this. 
The main idea is as follows. We assume strong log-concavity (e.g. Gaussianity) of the STP of $X_{t,r}$. Thus $D^i(X_{t,r})$ will be strongly convex with a unique minimizer at $\fr$. But $E_{Y_t}(X_t)$ (and so $E_{Y_t}$ as a function of $X_{t,r}$) can have multiple minimizers since OL can be multimodal. Assume that $E_{Y_t}(X_{t,s}^i, X_{t,r})$ is locally convex in the neighborhood of $\fr$ (this will hold if $\fr$ is close enough to any of its minimizers). Denote this region by $\R_{LC}$. Thus, inside $\R_{LC}$, $L^i$ will be strongly convex and hence it will have at most one minimizer. We show that if $\max_p |[\nabla D]_p|$ is large enough outside $\R_{LC}$ (the spread of the STP of $X_{t,r}$ is small enough), $L^i$ will have no stationary points (and hence no minimizers) outside $\R_{LC}$ or on its boundary.

This idea leads to Theorem \ref{unimodthm} below. Its first condition ensures strong convexity of $D^i$ everywhere. The second one ensures that $\R_{LC}$ exists. The third one ensures that $\exists$ an $\eps_0>0$, s.t. at all points in $\R_{LC}^c$ (complement of $\R_{LC}$),  $\max_p |[\nabla L^i]_p| > \eps_0$ (i.e. $L^i$ has no stationary points in $\R_{LC}^c$). 


\begin{theorem}
$\pss(X_{t,r})$ is unimodal with the unique mode lying inside $\R_{LC}$ if Assumption \ref{hmmass} and the following hold:%
\ben
\item The STP of $X_{t,r}$, $p(X_{t,r}|X_{t-1}^i,X_{t,s}^i)$, is strongly log-concave. Its unique mode is denoted by $\fr$. 
\label{slc}

\item The $-\log$ of OL given $X_{t,s}^i$, $E_{Y_t}(X_{t,s}^i,X_{t,r})$ is twice continuously differentiable almost everywhere and is locally convex in the neighborhood of $\fr$. 
Let $\R_{LC}  \subseteq \re^{M_r}$ denote the largest convex region in the neighborhood of $\fr$ where $\nabla_{X_{t,r}}^2 E_{Y_t}(X_{t,s}^i, X_{t,r}) \ge 0$ ($E_{Y_t}$ as a function of $X_{t,r}$ is locally convex).
\label{close}

\item There exists an $\eps_0>0$ such that
\bea
 \inf_{X_{t,r} \in  \cap_{p=1}^{M_r} (\A_{p} \cup \Z_{p})} \max_{p=1,\dots M_r} [ \gamma_p(X_{t,r})] > 1  \ \ 
\label{defcond}
\eea
where
\bea
\label{defgamma}
\gamma_p(X_{t,r}) \sdefn \left\{ \begin{array}{cc}
                    \frac{|[\nabla D^i]_p|}{\eps_0 +|[\nabla E_{Y_t}]_p|}, & \ \ if \ \ X_{t,r} \in \A_{p} \\ \\
                     \frac{|[\nabla D^i]_p|}{\eps_0 -|[\nabla E_{Y_t}]_p|} ,   & \ \ if  \ \ X_{t,r} \in \Z_{p}
                     \end{array}
                     \right.
\eea
\bea
\A_{p} \sdefn \{ X_{t,r} \in \R_{LC}^c: [\nabla D^i]_p.[\nabla E_{Y_t}]_p <  0\} \nn \\
\Z_{p} \sdefn \{ X_{t,r} \in \R_{LC}^c:  \ \nn \\
\label{defAZ}
         && [\nabla E_{Y_t}]_p. [\nabla D^i]_p \ge 0 \  \& \ |[\nabla E_{Y_t}]_p| < \eps_0 \} \ \ \  \ \ \ \\
\nabla E_{Y_t} \sdefn \nabla_{X_{t,r}}E_{Y_t}(X_{t,s}^i, X_{t,r}) \nn \\
\nabla D^i  \sdefn \nabla_{X_{t,r}} D^i(X_{t,r})
\label{defED}
\eea
\label{deltacond}
\een
\label{unimodthm}
\end{theorem}
\newcommand{\nep}{[\nabla E_{Y_t}]_p}
\newcommand{\ndp}{[\nabla D]_p} 
\proof{ In the proof, $\nabla$ is used to denote $\nabla_{X_{t,r}}$. Also, we remove the superscripts from $L^i$ and $D^i$.
$\pss(X_{t,r})$ will be unimodal iff  $L$ defined in (\ref{defL}) has at most one minimizer. We obtain sufficient conditions for this.
  Condition \ref{slc}) ensures that $D$ is strongly convex everywhere with a unique minimizer at $\fr$. Condition \ref{close}) ensures that $\R_{LC}$ exists. By definition of $\R_{LC}$, $E_{Y_t}$ is convex inside it. Thus the first two conditions ensure that $L$ is strongly convex inside $\R_{LC}$. So it has at most one minimizer inside $\R_{LC}$.

We now show that if condition \ref{deltacond}) also holds, $L$ will have no stationary points (and hence no minimizers) in $\R_{LC}^c$ or on its boundary. A sufficient condition for this is: $\exists \ \eps_0 > 0$ s.t.%
\bea
\max_p |[\nabla L]_p| > \eps_0, \ \forall X_{t,r} \in \R_{LC}^c
\label{cond}
\eea
We show that condition \ref{deltacond}) is sufficient to ensure (\ref{cond}).
Note that $\nabla L = \nabla E_{Y_t} + \nabla D$. In the regions where for at least one $p$, $\nep . \ndp \ge 0$ (have same sign) and $|\nep| > \eps_0$, condition (\ref{cond}) will always hold. Thus we only need to worry about regions where, for all $p$, either $\nep . \ndp < 0$ or $\nep . \ndp \ge 0$ but $|\nep| < \eps_0$. This is the region
\bea
\cap_{p=1}^{M_r} (\A_{p} \cup \Z_{p}) \defn \reg,  \mbox{~~$\A_p$, $\Z_p$ defined in (\ref{defAZ})}
\eea
Now, $D$ only has one stationary point which is $\fr$ and it lies inside $\R_{LC}$ (by definition of $\R_{LC}$), and none in $\R_{LC}^c$. Thus $\nabla D \neq 0$ in $\R_{LC}^c$ and, in particular, inside $\reg \subset \R_{LC}^c$. Thus if we can find a condition which ensures that, for all points in $\reg$,  for at least one $p$,  $[\nabla L]_p$ ``follows the sign of $\ndp$" (i.e. $[\nabla L]_p>\eps_0$ where $\ndp>0$ and $[\nabla L]_p < -\eps_0$ where $[\nabla D]_p < 0$), we will be done.

We first find the required condition for a given $p$ and a point $X_{t,r} \in \reg$. For any $p$, if $X_{t,r} \in \reg$, then it either belongs to $\A_p$ or belongs to $\Z_p$. If $X_{t,r} \in \A_p$, $|[\nabla L]_p| > \eps_0$ if
\bea
\frac{|\ndp|}{\eps_0 +|\nep|} > 1
\label{condAp}
\eea
This is obtained by combining the conditions for the case $\ndp>0$ and the case $\ndp < 0$. Proceeding in a similar fashion, if $X_{t,r} \in \Z_p$, $|[\nabla L]_p| > \eps_0$ if
\bea
\frac{|\ndp|}{\eps_0 -|\nep|} > 1
\label{condZp}
\eea
Inequalities (\ref{condAp}) and (\ref{condZp}) can be combined and rewritten as $\gamma_p(X_{t,r}) - 1 > 0$ where $\gamma_p$ is defined in (\ref{defgamma}).
For (\ref{cond}) to hold, we need $|[\nabla L]_p| > \eps_0$ for at least one $p$, for all $X_{t,r} \in \reg$. This will happen if $\inf_{X_{t,r} \in \reg} \max_p  \gamma_p(X_{t,r})  >1$. But this is condition \ref{deltacond}. Thus condition \ref{deltacond}) implies that $L$ has no minimizers in $\R_{LC}^c$.
Thus if conditions \ref{slc}), \ref{close}) and \ref{deltacond}) of the theorem hold, $L$ has at most one minimizer which lies inside $\R_{LC}$. Thus  $\pss(X_{t,r})$ has a unique mode which lies inside $\R_{LC}$, i.e. it is unimodal.
} $\blacksquare$

The most common example of a strongly log-concave pdf is a Gaussian. When the STP of $X_{t,r}$ is Gaussian with  mean (= mode) $\fr$, the above result can be further simplified to get an upper bound on the eigenvalues of its covariance matrix. First consider the case when the covariance is diagonal, denoted $\Delta_r$. In this case, $D^i(X_{t,r}) = \sum_p \frac{([X_{t,r} - \fr]_p)^2}{2\Delta_{r,p}}$ and so $[\nabla D^i ]_p = \frac{[X_{t,r} - \fr]_p}{\Delta_{r,p}}$. By substituting this in condition \ref{deltacond}), it is easy to see that we get the following simplified condition:
\bea
 \inf_{X_{t,r} \in  \cap_{p=1}^{M_r} (\A_{p} \cup \Z_{p})} \ \max_{p} [ \gamma^{num}_p(X_{t,r}) - \Delta_{r,p}] > 0  \ \
\label{defcond2}
\eea
\bea
\label{defgamma2}
\gamma^{num}_p(X_{t,r}) \sdefn \left\{ \begin{array}{cc}
                    \frac{|[X_{t,r} - \fr]_p|}{\eps_0 +|[\nabla E_{Y_t}]_p|}, & \ \ if \ \ X_{t,r} \in \A_{p} \\ \\
                     \frac{|[X_{t,r} - \fr]_p}{\eps_0 -|[\nabla E_{Y_t}]_p|} ,   & \ \ if  \ \ X_{t,r} \in \Z_{p}
                     \end{array}
                     \right.
\eea
\bea
\A_{p} \sdefn \{ X_{t,r} \in \R_{LC}^c: [X_{t,r} - \fr]_p.[\nabla E_{Y_t}]_p <  0\} \nn \\
\Z_{p} \sdefn \{ X_{t,r} \in \R_{LC}^c:  \ \nn \\
         && [\nabla E_{Y_t}]_p. [X_{t,r} - \fr]_p \ge 0 \  \& \ |[\nabla E_{Y_t}]_p| < \eps_0 \} \ \ \
\label{defAZ2}
\eea
Also, since $\max_p[g_1(p)-g_2(p)] \ge \max_p g_1(p)-\max_p g_2(p)$ for any two functions, $g_1$, $g_2$, a sufficient condition for (\ref{defcond2}) is
\bea
\max_p  \Delta_{r,p} < \inf_{X_{t,r} \in  \cap_{p=1}^{M_r} (\A_{p} \cup \Z_{p})} \ \max_{p} [ \gamma^{num}_p(X_{t,r})] \defn \Delta^* \ \ \ \ \
\label{suff}
\eea
Thus, we have the following corollary.
\begin{corollary}
When the STP of $X_{t,r}$ is Gaussian with mean $\fr$ and {\em diagonal covariance, $\Delta_r$}, $\pss(X_{t,r})$ is unimodal if (a) condition \ref{close}) of Theorem \ref{unimodthm} holds and (b) there exists an $\eps_0>0$ s.t. (\ref{defcond2}) holds with $\gamma^{num}_p$ defined in (\ref{defgamma2}) and  $\A_p, \Z_p$ defined in (\ref{defAZ2}). A sufficient condition for (\ref{defcond2}) is (\ref{suff}).%
\label{diagcov}
\end{corollary}

Now consider the case when the STP of $X_{t,r}$ is Gaussian with non-diagonal covariance, $\Sigma_r = U \Delta_r U^T$. Define $\tilde{X}_{t,r} = U^T X_{t,r}$. Since $\tilde{X}_{t,r}$ is a one-to-one and linear function of $X_{t,r}$, it is easy to see that $\pss(X_{t,r})$ is unimodal iff $\pss(\tilde{X}_{t,r}) \defn p(\tilde{X}_{t,r}|X_{t-1}^i,X_{t,s}^i,Y_t)$ is unimodal. The STP of $\tilde{X}_{t,r}$ is $\n(U^T \fr, \Delta_r)$. Also, its OL is $p(Y_t|X_{t,s}^i, U \tilde{X}_{t,r})$. Define $\tilde{E}_{Y_t}(\tilde{X}_{t,r}) \defn E_{Y_t}(U\tilde{X}_{t,r})$.
\begin{corollary}
When the STP of $X_{t,r}$ is Gaussian with mean $\fr$ and {\em non-diagonal covariance, $\Sigma_r = U \Delta_r U^T$}, $\pss(X_{t,r})$ is unimodal  if the conditions of Corollary \ref{diagcov} hold with $E_{Y_t}$ replaced by $\tilde{E}_{Y_t}$; $\fr$ replaced by $U^T\fr$ and $X_{t,r}$ replaced by $\tilde{X}_{t,r}$ everywhere.
\label{nondiagcov}
\end{corollary}

To summarize the above discussion, $\pss$ is unimodal if
\ben
\item The STP of  $X_{t,r}$ is strongly log-concave (e.g. Gaussian),
\item The mode of the STP of $X_{t,r}$ is ``close enough" to a mode of [OL given $X_{t,s}^i$], so that condition \ref{close}) of Theorem \ref{unimodthm} holds. Denote this mode by $X_{r}^*$.

\item The maximum spread of the STP of  $X_{t,r}$ is ``small enough" to ensure that condition \ref{deltacond}) of Theorem \ref{unimodthm} holds.
In the Gaussian STP case, this translates to the maximum eigenvalue of its covariance being smaller than $\Delta^*$, defined in (\ref{suff}).  $\Delta^*$ itself is directly proportional to the distance of  $X_{r}^*$ to the next nearest mode of [OL given $X_{t,s}^i$] and inversely proportional to its strength. 
\een
{\em The last two conditions above automatically hold if [OL given $X_{t,s}^i$] is strongly log-concave} ($\R_{LC}^c$ is empty and so $\Delta^* = \infty$).

\section{A Generic State Space Model for LDSS}
\label{ldssmodel}
For many problems, and, in particular, for many large dimensional state space (LDSS) problems, the state space model can be expressed as follows with $X_t = [C_t, v_t]$ (a generalization of the constant velocity motion model):
\bea
Y_t \se h_{C,w}(C_t, w_t), \ \ w_t \sim p_w(.) \nn \\
C_t \se C_{t-1} + g_{C_{t-1}}( B v_t ), \ B \defn B(C_{t-1})    \nn \\
v_t \se f_v( v_{t-1} ) + \nu_t, \  \nu_t \sim \n(0,\Delta_\nu), \ \Delta_\nu \ \mbox{diagonal} \ \ \ 
\label{ldssmod}
\eea
The noises $\nu_t$, $w_t$ are independent of each other and over time. 
If $h_{C,w}$ is one-to-one as a function of $w_t$, and its inverse is denoted by $g(C_t,Y_t)$, the OL can be written as
\bea
p(Y_t|C_t) = p_w( g(C_t,Y_t) )
\eea
Then its $-\log$, $E_{Y_t}(C_t) = -\log p_w( g(C_t,Y_t) )$. In certain problems, it is easier to directly specify  $p(Y_t|C_t) = \beta \exp[-E_{Y_t}(C_t)]$.
In the above model, $C_t$ denotes the LDSS quantity of interest, for e.g. it may denote the $M$ contour point locations or it may denote temperature (or any other physical quantity) at $M$ sensor nodes. The quantity $V_t \defn B v_t$ often denotes the time ``derivative" of $C_t$  and is assumed to follow a first order Markov model. If $C_t$ belongs to a smooth manifold $\spc$, then $V_t$ belongs to the tangent space to $\spc$ at $C_t$. $g_C(V)$ denotes the mapping from the tangent space at $C$ to $\spc$, while if $\spc$ is a vector space, then $g_C(V) \equiv V$. In this work, we only study the vector space case. We develop the same ideas for the space of contours (a smooth manifold) in \cite{cdc06journal}.
Related work on defining AR models for smooth manifolds is \cite{armanifold}.

Note that in the above model, the {\em system noise dimension (and hence the importance sampling dimension) is $M = \dim(\nu_t) = \dim(v_t)$}, and not $2M$, and this is what governs the number of particles required for a given accuracy.

We discuss some LDSS examples below.

\begin{example}[Temperature Tracking]
Consider tracking temperature at $M$ locations using a network of sensors. Here $\spc$ is a vector space and so $g_C(V) \equiv V$. Let $C_{t,p}$ denotes temperature at location $p$, $p=1,\dots M$ and $V_{t,p}$ denote the first derivative of temperature at node $p$. $V_t$ is assumed to be zero mean and its dynamics can be modeled by a linear Gauss Markov model (as also in \cite{mouraRC}), i.e.
\bea
C_t \se C_{t-1} +  V_t, \ 
V_t = A_V V_{t-1} + n_t, \ n_t \sim \n(0,\Sigma_n) \ \ \
\label{ldsstemp0}
\eea
Since $V_t$ is usually spatially correlated, $\Sigma_n$ may not be diagonal. Let the eigenvalue decomposition of $\Sigma_n$ is $\Sigma_n = B \Delta_\nu B^T$. Define $v_t \defn B^T V_t$, $\nu_t \defn B^T n_t$, $f_v(v) \equiv B^T A_V B v$ and $g_C(V) \equiv V$. For simplicity, we use $A_V = aI$ and so $f_v(v) \equiv B^T A_V B v = a v$. With $f_v(v)=a v$, $B$ is also the eigenvector matrix of the covariance of $V_t$. Then (\ref{ldsstemp0}) can be rewritten in the form (\ref{ldssmod}) as
\bea
C_t \se C_{t-1} + B v_t \nn \\
v_t \se a v_{t-1} + \nu_t, \ \nu_t \sim \n(0,\Delta_\nu)
\label{ldsstemp}
\eea
%
 Temperature at each node, $p$, is measured using $J$ ($J=1$ or 2) sensors that have failure probabilities $\alpha_p^{(j)}, j=1,2$. Note that there may actually be two sensors at a node, or two nearby sensors can be combined and treated as one ``node" for tracking purposes. Failure of the $J M$ sensors is assumed to be independent of each other and over time. If a sensor is working, the observation, $Y_{t,p}^{(j)}$, is the actual temperature, $C_{t,p}$, or some function of it, $h_p(C_{t,p})$, plus Gaussian noise with small variance, $\sigma_{obs,p}^2$ (independent of noise in other sensors and at other times). If the sensor fails, $Y_{t,p}^{(j)}$ is either independent of, or weakly dependent on $C_{t,p}$ (e.g. large variance Gaussian about $C_{t,p}$). An alternative failure model is $Y_{t,p}^{(j)}$ being some different function, $h_p^f$, of $C_{t,p}$ plus noise. 
In all the above cases, the OL can be written as
\bea
p(Y_t|C_t) \se \prod_{p=1}^M  \prod_{j=1}^J p(Y_{t,p}^{(j)}|C_{t,p}), \ where \nn \\ 
p(Y_{t,p}^{(j)}|C_{t,p}) \se (1-\alpha_p^{(j)}) \ \n(Y_{t,p}^{(j)}; h_p(C_{t,p}), \sigma_{obs,p}^2)   \nn \\
 &&  + \ \alpha_p^{(j)}  \  p_f(Y_{t,p}^{(j)}|C_{t,p}) 
\label{sensorobsmod}
\eea
We simulated two types of sensors $h_p(C_{t,p}) = C_{t,p}$ (linear) and $h_p(C_{t,p}) = C_{t,p}^2$ (squared). Note that a squared sensor is an extreme example of possible sensing nonlinearities. 
 First consider $J=1$ (one sensor per node), $h_p(C_{t,p}) = C_{t,p}, \forall p$ (all linear sensors), and $p_f(Y_{t,p}^{(j)}|C_{t,p}) = p_f(Y_{t,p}^{(j)})$ (when the sensor fails, the observation is independent of the true temperature). In this case, each OL term is a raised Gaussian (heavy-tailed) as a function of $C_{t,p}$ and so it is not strongly log-concave. For a given $p$, $p^*(C_{t,p})$ will be multimodal when $Y_{t,p}^{(1)}$ is ``far" from the predicted temperature at this node and the STP is not narrow enough. This happens with high probability (w.h.p.) whenever the sensor fails. See Fig. \ref{fails2}. A similar model is also used in modeling clutter \cite{condensation,asilomar}.

Now consider $J=2$, all linear sensors and $p_f(Y_{t,p}^{(j)}|C_{t,p}) = p_f(Y_{t,p}^{(j)})$. Whenever one or both sensors at a node $p_0$ fail, the observations $Y_{t,p_0}^{(1)}, Y_{t,p_0}^{(2)}$ will be ``far" compared to $\sigma_{obs}^2$ w.h.p. In this case, the OL will be bimodal as a function of $C_{t,p_0}$ since $p(Y_{t,p}|C_{t,p})$ can be written as a sum of four terms: a product of Gaussians term (which is negligible), plus $K_1 + K_2\n(Y_{t,p_0}^{(1)}; C_{t,p_0}, \sigma_{obs}^2) + K_3 \n(Y_{t,p_0}^{(2)}; C_{t,p_0}, \sigma_{obs}^2)$ where $K_1, K_2, K_3$ are constants w.r.t. $C_{t}$. This is bimodal since the modes of the two Gaussians, $Y_{t,p_0}^{(1)}$, $Y_{t,p_0}^{(2)}$, are ``far". See Fig. \ref{fails}. If no sensor at a node fails, both observations will be ``close" w.h.p.. In this case all four terms have roughly the same mode, and thus the sum is unimodal. When $p_f(Y_{t,p}^{(j)}|C_{t,p})$ is  weakly dependent on $C_{t,p}$ (e.g. a large variance Gaussian), $K_1, K_2, K_3$ are not constants but are slowly varying functions of $C_t$. A similar argument applies there as well.

A squared sensor results in a bimodal OL whenever $Y_{t,p}^{(j)}$ is significantly positive. Squared sensor is one example of a many-to-one measurement function. Other examples include bearings-only tracking \cite{gordon} and illumination tracking \cite{kalecvpr,kaleicassp}.

%
%
%
\label{temptrack}
\end{example}
\begin{example}[Illumination/Motion Tracking] 
The illumination and motion tracking model of \cite{kalecvpr,kaleicassp} can be rewritten in the form (\ref{ldssmod}). In this case, the OL is often multimodal since the observation (image intensity) is a many-to-one function of the state (illumination, motion), but conditioned on motion, it is often unimodal. The STP of illumination is quite narrow.
\label{illumtrack}
\end{example}
\begin{example}[Contour Tracking, Landmark Shape Tracking]
Two non-Euclidean space examples of the LDSS model (\ref{ldssmod}) are (a) the contour tracking problems given in \cite{cdc06journal,pami07,wilsky} and (b) the landmark shape tracking problem of \cite{asilomar,condensation}.
\label{conttrack}
\end{example}



\begin{figure*}[t!]
\begin{center}
\subfigure[Region $\R_{LC}$ \& point $\fr$]
{\label{RLCfig}
\epsfig{file = 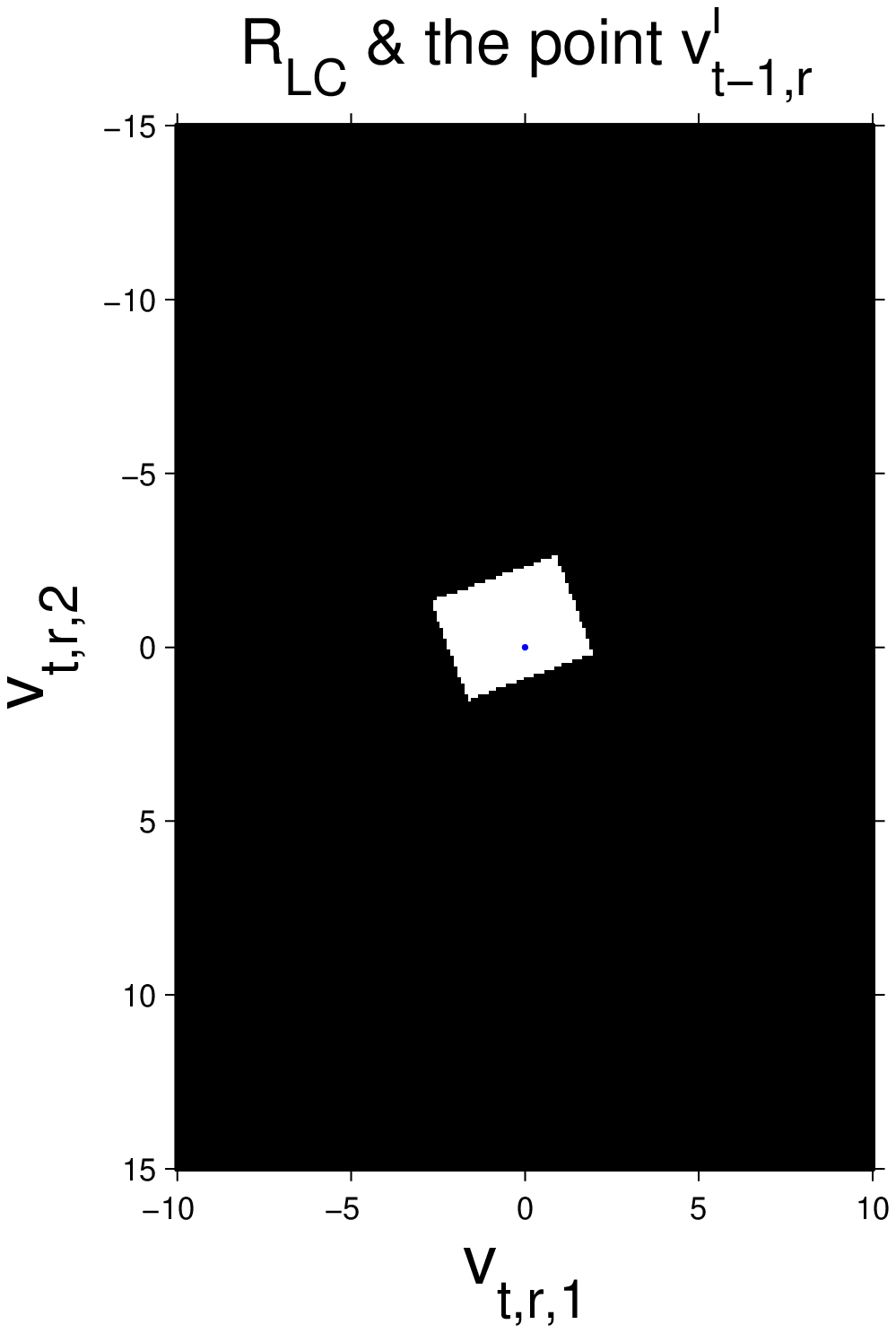, width=3cm,height=3cm}
}
\subfigure[$\A_1 \cap \A_2$]
{\label{A1A2}
\epsfig{file = 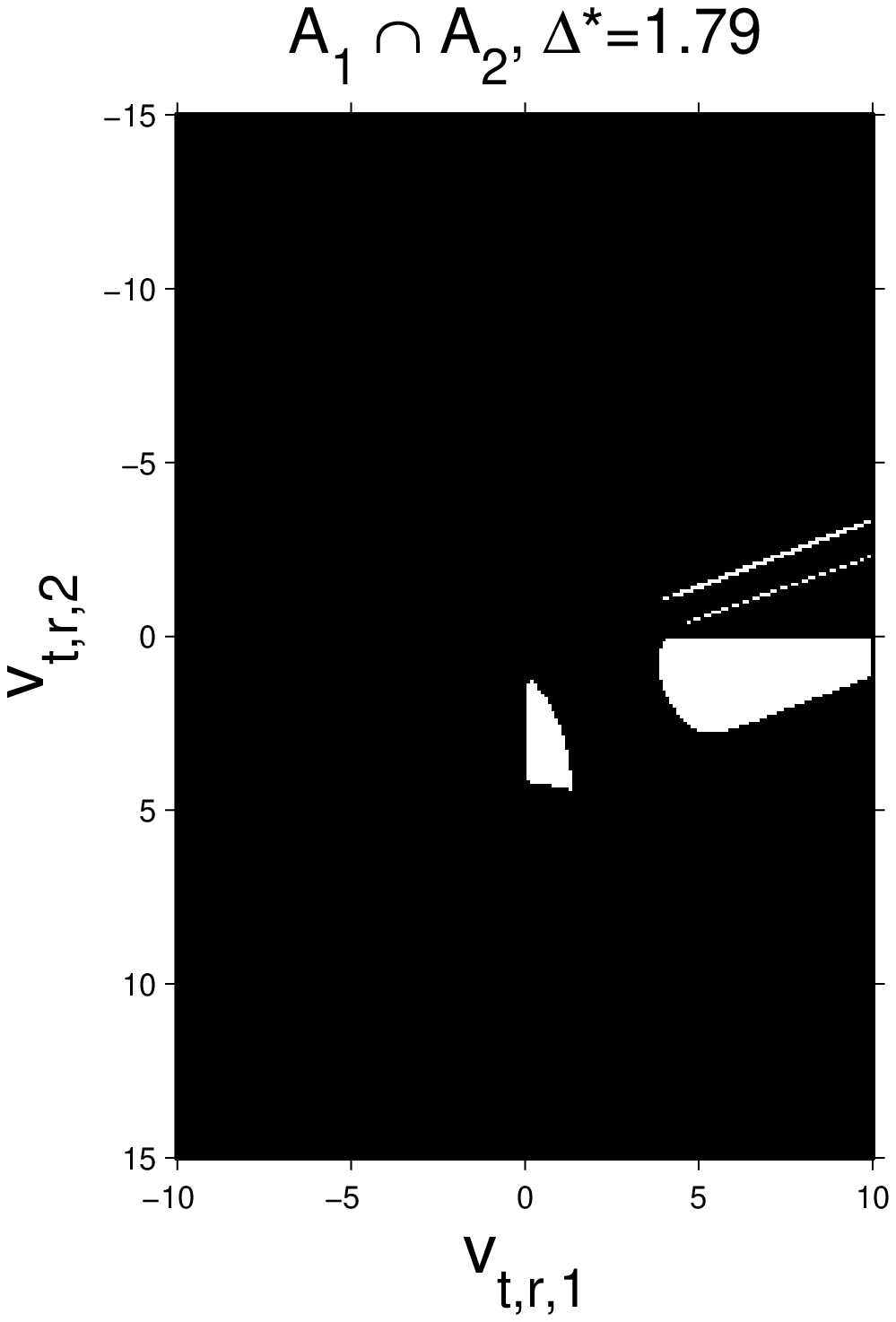, width=3cm,height=3cm}
}
\subfigure[$\Z_1 \cap \A_2$]
{\label{Z1A2}
\epsfig{file = 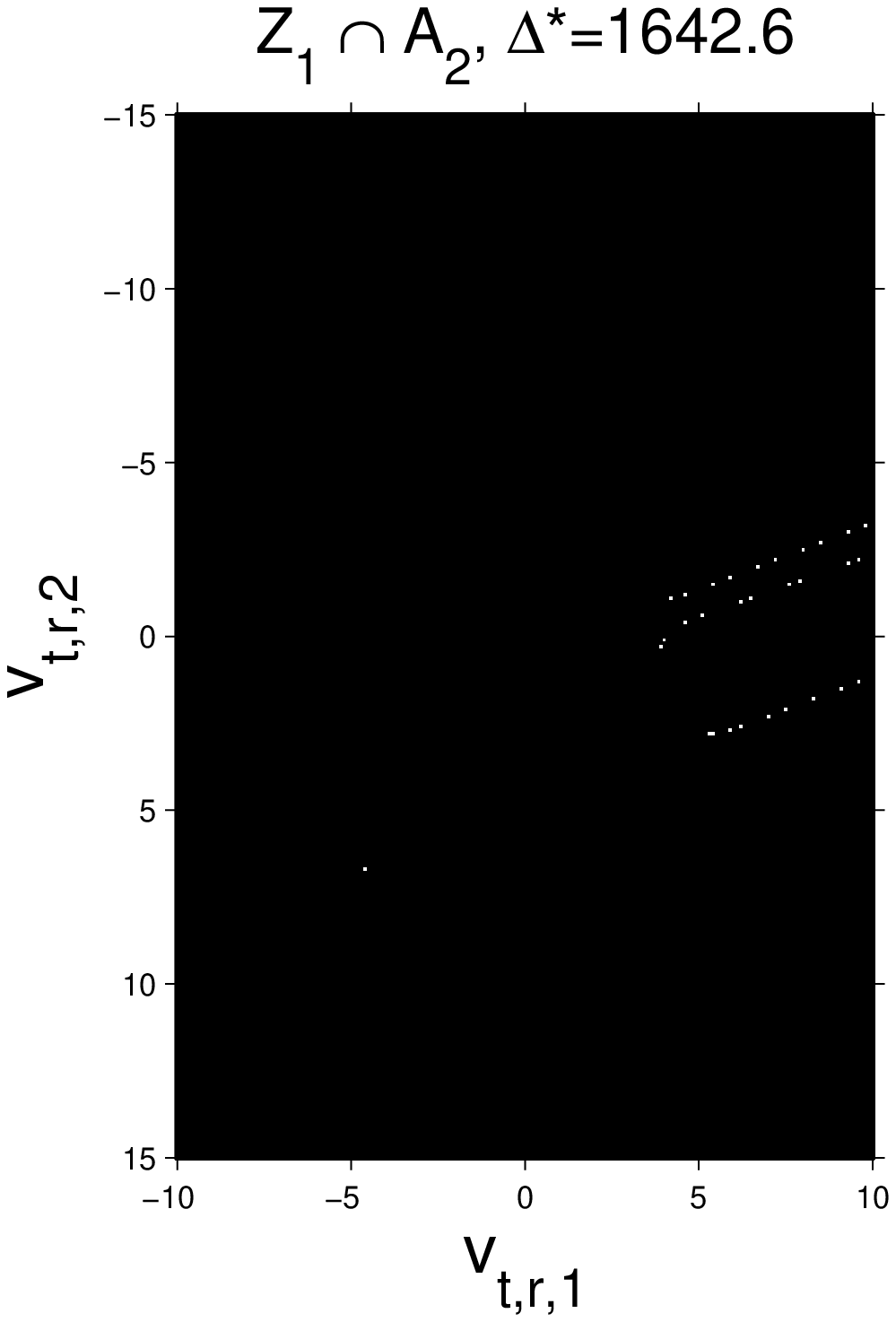, width=3cm,height=3cm}
}
\subfigure[$\Z_2 \cap \A_1$]
{\label{Z2A1}
\epsfig{file = 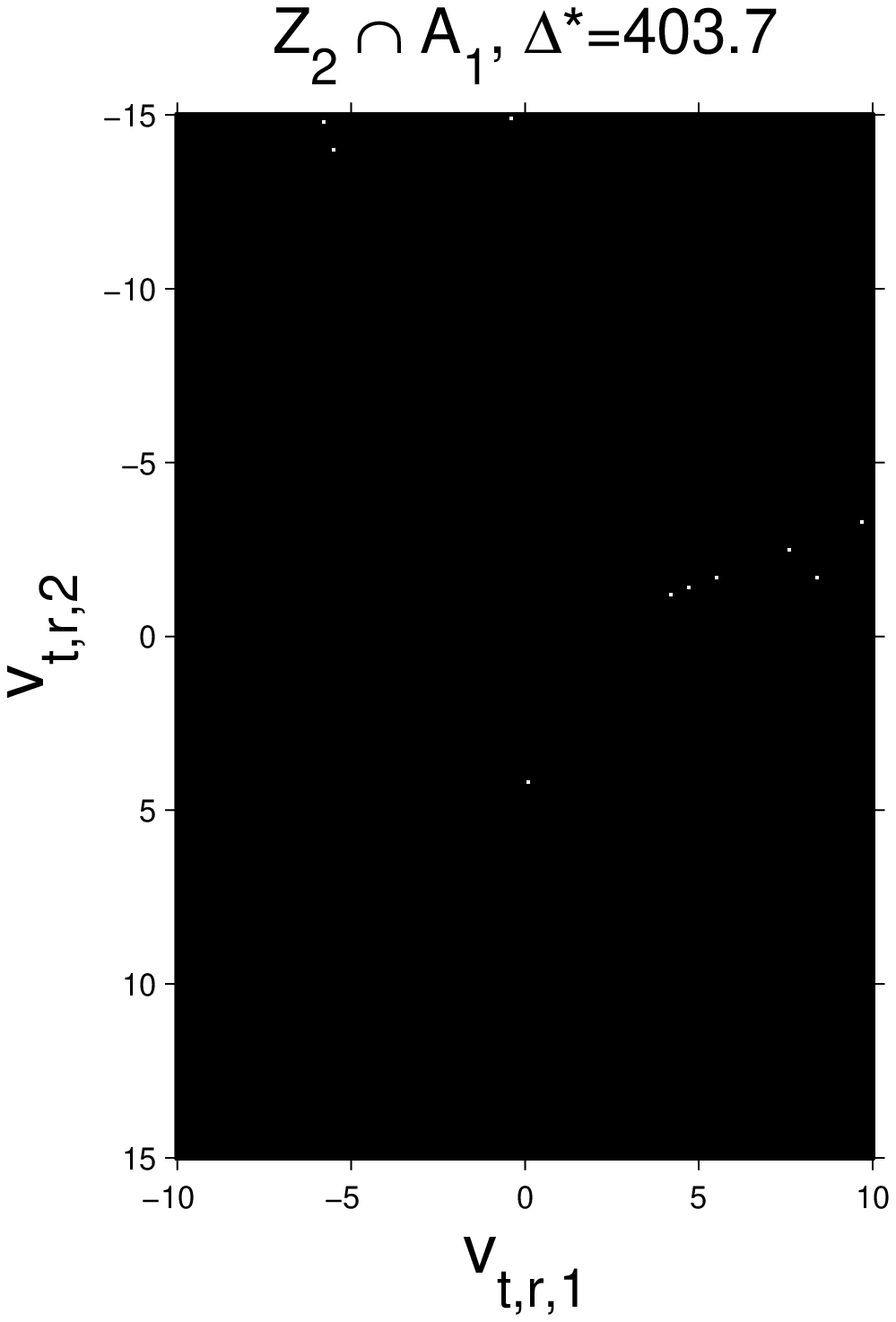, width=3cm,height=3cm}
}
\subfigure[$\Z_1 \cap \Z_2$]
{\label{Z2Z1}
\epsfig{file = 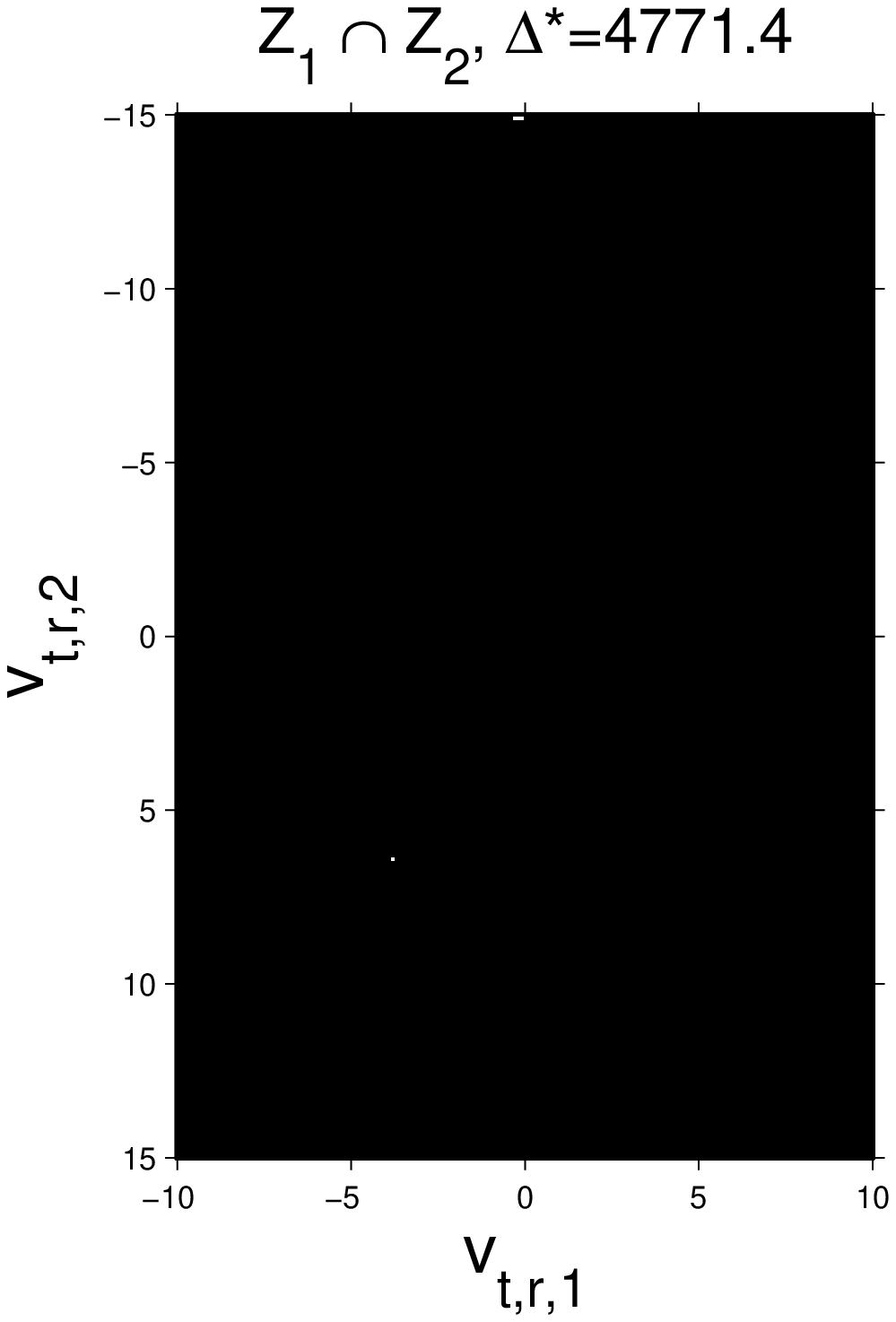, width=3cm,height=3cm}
}
\\
\subfigure[$E_{Y_t}(v_{t,r})$]
{\label{Efig}
}
\subfigure[$(\nabla E_{Y_t})_p = 0,  p=1,2$]
{\label{nablaEfig}
\epsfig{file = 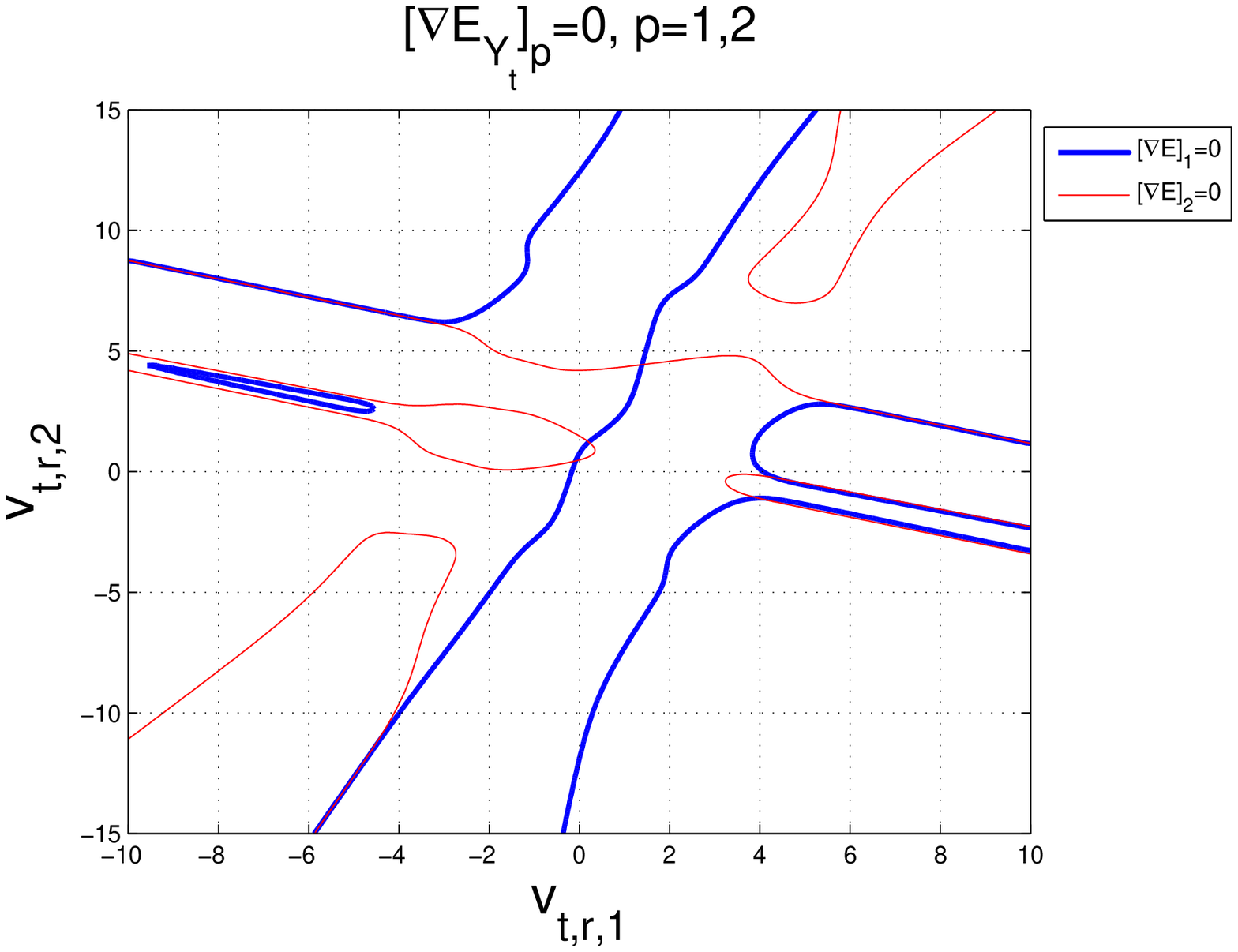, width=4cm,height=3cm}
}
\subfigure[$(\nabla L)_p=0$, \ $\Delta_r = 0.9 \Delta^*$]
{\label{nablaLunimodfig}
\epsfig{file = 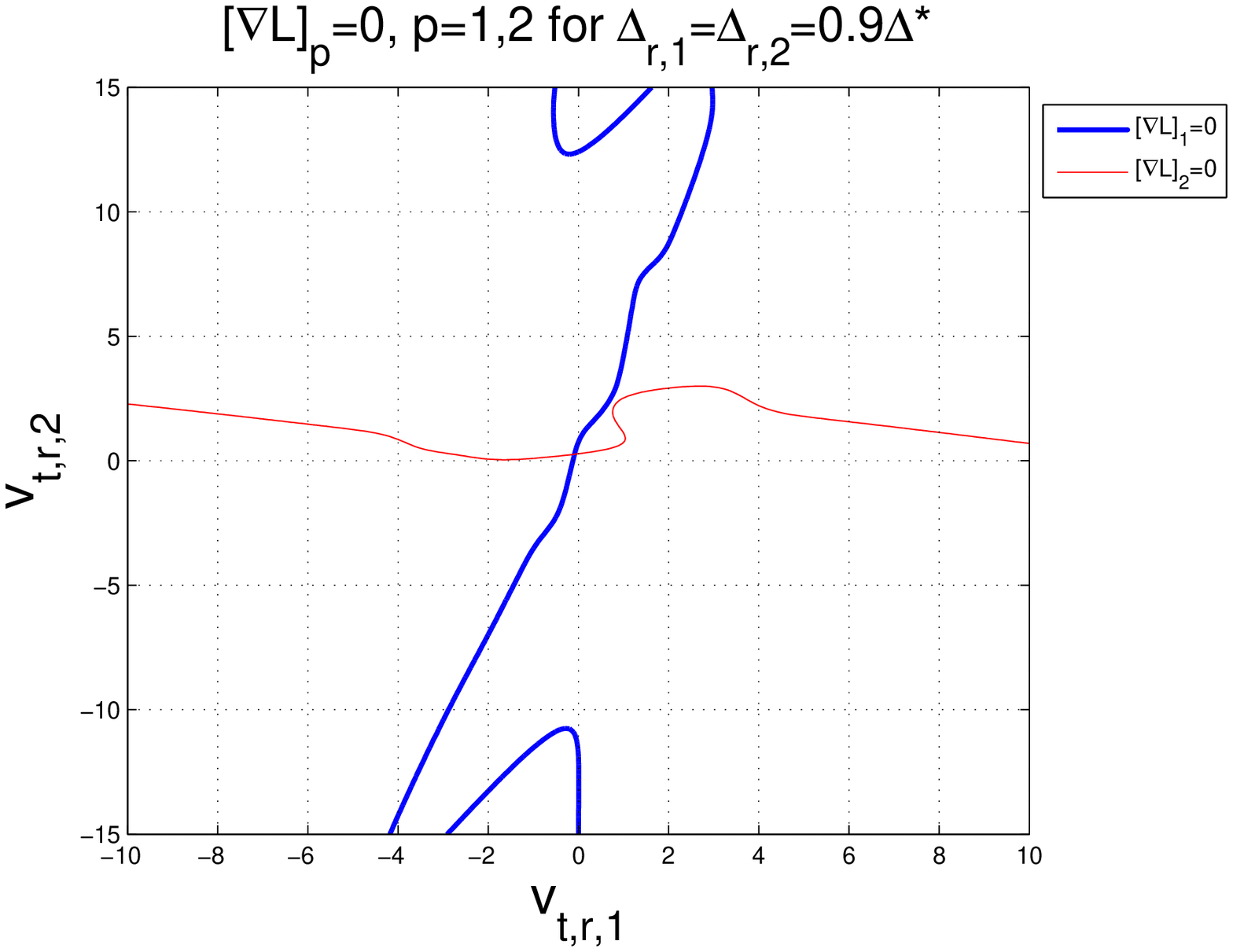, width=4cm,height=3cm}
}
\subfigure[$(\nabla L)_p=0$, \ $\Delta_r = 1.1 \Delta^*$]
{\label{nablaLmultimodfig}
\epsfig{file = 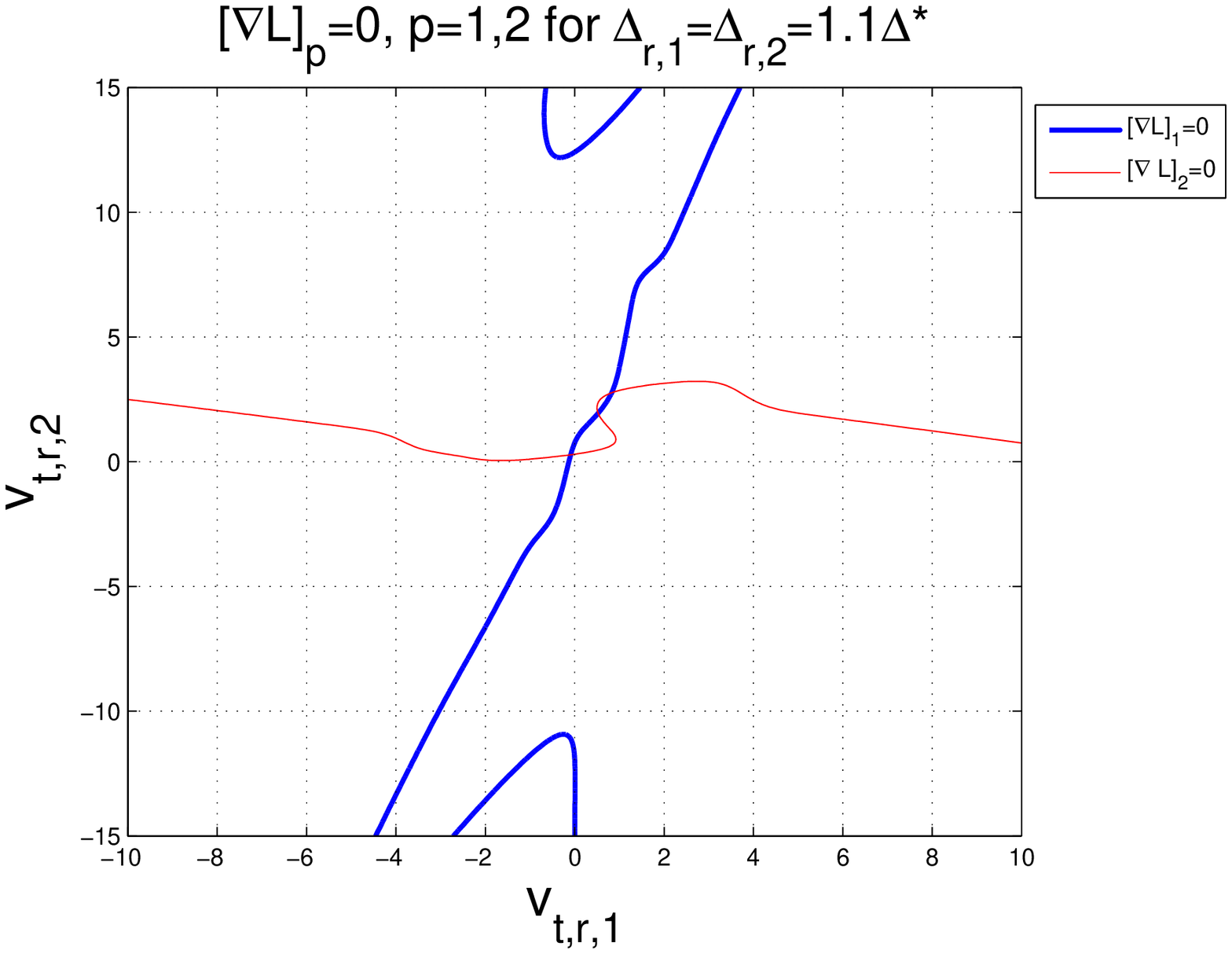, width=4cm,height=3cm}
}
\end{center}
\vspace{-0.1in}
\caption{{\small Computing $\Delta^*$ for Example \ref{example1}. 
We used $\alpha^{(1)}=[0.1, 0.1, 0.1]$, $\alpha^{(2)} = [0.4, 0.4, 0.4]$, $p_f(Y_{t,p}^{(j)})= Unif(-10,10), j=1,2, \ \forall p$, $\sigma_{obs}^2 = [1, 1, 1]$, $\Delta_{\nu,1} =5.4$, $B = [-0.27, 0.96, -0.02]'; [0.33,  0.11,  0.94]'; [0.90, 0.24   -0.35]'$ (we use MATLAB notation). Also, $C_{t-1}^i=[ 0, 0, 0]'$, $v_{t-1,r}^i=[0 \ 0]'$, $v_{t-1,s}^i=0$,  $Y_{t,1}^{(1,2)} = [5.36,   0.59], \ Y_{t,2}^{(1,2)} = [ -2.25,  -1.60]  \ Y_{t,3}^{(1,2)} = [-0.68, 0.35]$ and $v_{t,s}^i = -3.2$ (simulated from $\n(0, \Delta_{\nu,1})$).
Fig. \ref{RLCfig}: region $\R_{LC}$, and the point $\fr=v_{t-1,r}^i$ which lies inside it.
Fig. \ref{A1A2}, \ref{Z1A2}, \ref{Z2A1}, \ref{Z2Z1}: the regions, $\A_{1} \cap \A_{2}$, $\Z_{1} \cap \A_{2}$, $\Z_{2} \cap \A_{1}$ and $\Z_{1} \cap \Z_{2}$, along with the computed minimum value of $\max_p \gamma_p(v_{t,r})$ in the 4 regions (1.79, 1642.6, 403.7, 4771.4). The final value of $\Delta^*$ is the minimum of these four values, i.e.  $\Delta^* =1.79$.
Fig \ref{Efig}: mesh plot of $E_{Y_t}$ as a function of $v_{t,r}$. Note the 2 dominant modes.
Fig \ref{nablaEfig}: contours of $[\nabla E_{Y_t}]_1=0$ and $[\nabla E_{Y_t}]_2=0$  (obtained using the contour command to find the zero level set of $[\nabla E_{Y_t}]_j,j=1,2$). The contours have many points of intersection  (points where $\nabla E_{Y_t} = 0$), i.e. many stationary points.
Fig \ref{nablaLunimodfig}: contours of $[\nabla L]_1=0$ and $[\nabla L]_2=0$ for $L$ computed with $\Delta_{\nu,2} = \Delta_{\nu,3} = 0.9 \Delta^*$. The contours have only one point of intersection  which is a minimum. 
Fig \ref{nablaLmultimodfig}: contours of of $[\nabla L]_j=0,j=1,2$ for $\Delta_{\nu,2} = \Delta_{\nu,3} = 1.1 \Delta^*$. There are 3 intersection points (2 are minima).
}}
\label{computeDelta}
\end{figure*}

\section{Choosing the ``Multimodal States" for LDSS}
\label{effbasis}
In Sec.\ref{unimodldssres} below, we apply Theorem \ref{unimodthm} to the generic LDSS model, (\ref{ldssmod}), and show an example of verifying its conditions. Practical ways to select $X_{t,s}$ are given in Sec.\ref{chooseKunimod}.

\subsection{Unimodality Result for LDSS model} 
\label{unimodldssres}
Consider a model of the form (\ref{ldssmod}) with $g_C(V) \equiv V$. Assume that $v_t$ can be partitioned into $v_t = [v_{t,s};  v_{t,r}]$ where $v_{t,s}$ denotes the temperature change coefficients along the ``multimodal" directions of the state space and $v_{t,r}$ denotes the rest. Thus, $X_{t,s} = v_{t,s}$ and $X_{t,r} = [v_{t,r}, C_t]$.
Similarly partition $B = [B_s, B_r]$, $\Delta_\nu = diag(\Delta_{\nu,s}, \Delta_{\nu,r})$ and $\nu_t = [\nu_{t,s}; \nu_{t,r}]$. We discuss how to choose $v_{t,s}$ and $v_{t,r}$ in  Sec. \ref{chooseKunimod}.  The ``multimodal" dimension, $K = \dim(v_{t,s})$ and $M_r = M-K$.
Denote
\bea
\chat^i \sdefn C_{t-1}^i + B_s v_{t,s}^i, \
\fr \defn  f_{v,r}(v_{t-1}^i) \nn 
\eea
Then we have
\bea
&& \pss(v_{t,r}, C_{t}) \nn \\
&& = p(v_{t,r}, C_t | v_{t-1}^i, C_{t-1}^i, v_{t,s}^i, Y_t) \nn \\
&& = \zeta \ \n(v_{t,r}; \fr, \Delta_{\nu,r}) \ \delta(C_t - [\chat^i + B_r v_{t,r}]) \ p(Y_t|C_t)  \nn \\ 
&& = {\zeta  \n(v_{t,r}; \fr, \Delta_{\nu,r})  p(Y_t|\chat^i + B_r v_{t,r})}  \delta(C_t - [\chat^i + B_r v_{t,r}]) \nn \\
&& \defn \pss(v_{t,r}) \  \delta(C_t - [\chat^i + B_r v_{t,r}])
\eea
where $\delta$ denotes the Dirac delta function and $\zeta$ is a proportionality constant.
Since $C_t$ is a deterministic function of $C_{t-1}^i, v_{t,s}^i, v_{t,r}$, its pdf is a Dirac delta function (which is trivially unimodal at $\chat^i + B_r v_{t,r}$). Thus for the purpose of importance sampling, $X_{t,r} = v_{t,r}$ only, and we need conditions to ensure that $\pss(v_{t,r})$ is unimodal. In this case, $L^i(v_{t,r}) \defn -\log \pss(v_{t,r}) + \const$ becomes%
\bea
L^i(v_{t,r}) \defn E_{Y_t}(\chat^i + B_r v_{t,r}) + \sum_{p=1}^{M-K} \frac{ ( [v_{t,r} - \fr ]_p )^2 }{ 2 \Delta_{\nu,r,p} }
\label{defLm1}
\eea
Applying Corollary \ref{diagcov}  we get, 
\begin{corollary}
Consider model (\ref{ldssmod}) with $g_C(V) \equiv V$. Corollary \ref{diagcov} applies with the following substitutions:  
 $X_{t,s} \equiv v_{t,s}$, $X_{t,r} \equiv v_{t,r}$,  $M_r \equiv M-K$,
 $\Delta_{r,p} \equiv \Delta_{\nu,r,p}$,
 $\fr \equiv  f_{v,r}(v_{t-1}^i)$,  
 $\nabla E_{Y_t} \equiv B_r^T \nabla_C E (\chat^i + B_r v_{t,r})$,
 $\R_{LC} \subseteq \re^{M-K}$ is the largest convex region in the neighborhood of $\fr$ where $E_{Y_t}(\chat^i + B_r v_{t,r})$ is convex as a function of $v_{t,r}$.
\label{ldssvec}
\end{corollary}

We demonstrate how to verify the conditions of Corollary \ref{ldssvec} using a temperature tracking example.
We use numerical (finite difference) computations of gradients. Here $\eps_0$ needs to be chosen carefully, depending on the resolution of the discretization grid of $v_{t,r}$. It should be just large enough\footnote{If $\eps_0$ is too small, $\nep$ may transition from a value smaller than $-\eps_0$ to a value larger than $+\eps_0$ (or vice versa) over one grid point, and this region will not be included in $\Z_p$ (even if $\ndp$ has the same sign as $\nep$), thus getting a wrong (too large) value of $\Delta^*$. If $\eps_0$ is larger than required, the region $\Z_p$ may be bigger than needed, thus giving a smaller value $\Delta^*$ than what can actually be allowed.} so that one does not miss any stationary point of $E_{Y_t}$.

\begin{example}
Consider Example \ref{temptrack}. Assume that $M=3$ and OL follows (\ref{sensorobsmod}) with $h_p(C_{t,p}) = C_{t,p}$ (linear sensors) and $p_f(Y_{t,p}^{(j)}|C_{t,p}) = p_f(Y_{t,p}^{(j)})$. Also, let $a=1$. 
%
%
In Fig. \ref{computeDelta}, we demonstrate how to verify the conditions of Corollary \ref{ldssvec}. Let $K=1$, i.e $M_r=2$. Assume that $X_{t,s}=v_{t,s} = v_{t,1}$ and $v_{t,r} = v_{t,2:3}$. Assume a given value of $C_{t-1}^i$, $\fr$ and of $Y_t$ (given in the figure caption). Note that $Y_{t,1}^{(1)} =5.36$, $Y_{t,1}^{(2)}=0.59$ are ``far" compared to $\sigma_{obs,1}=1$ and hence the OL is multimodal. Fig. \ref{Efig} plots $E_{Y_t}(v_{t,r})$. Fig. \ref{nablaEfig} plots the contours of $[\nabla E_{Y_t}]_p=0, p=1,2$ (the points where the red and blue contours intersect are the stationary points of $E_{Y_t}$).

Verification of condition \ref{close} is shown in Fig. \ref{RLCfig}. Next, we show the steps for computing $\Delta^*$. For $M_r=2$, $\reg = \cap_{p=1}^2 (\A_p \cup \Z_p)$ is a subset of $\re^2$ and is a union of the 4 regions: $\A_{1} \cap \A_{2}$, $\Z_{1} \cap \A_{2}$, $\A_{1} \cap \Z_{2}$, $\Z_{1} \cap \Z_{2}$, shown in Fig \ref{A1A2}, \ref{Z1A2}, \ref{Z2A1}, \ref{Z2Z1}. The computed value of the minimum of $\max_p \gamma_p^{num}(v_{t,r})$ in each region is also given in the titles. The final $\Delta^*=1.79$ is the minimum of these 4 values. Contours of $[\nabla L^i]_1=0$ and of $[\nabla L^i]_2=0$ computed for $\Delta_{\nu,2}=\Delta_{\nu,3} = 0.9 \Delta^*$ and $1.1 \Delta^*$ are shown in Figs. \ref{nablaLunimodfig}, \ref{nablaLmultimodfig}. Notice that when $\Delta_{\nu,2}=\Delta_{\nu,3} = 0.9 \Delta^*$, they intersect at only one point i.e. $\nabla L^i = 0$ at only one point (one stationary point). When $\Delta_{\nu,2}=\Delta_{\nu,3} = 1.1 \Delta^*$, there are 3 stationary points (and 2 are minima).
%
\label{example1}
\end{example}

\subsection{Choosing the ``Multimodal" States, $X_{t,s}$}
\label{chooseKunimod} 
Corollary \ref{ldssvec} gives a unimodality condition that needs to be verified separately for each particle and each $Y_t$ at each $t$. An exact algorithm to do this would be to begin by checking at each $t$, for each $i$, if Theorem \ref{unimodthm} holds with $K=0$. Keep increasing $K$ and doing this until find a $K$ for which Corollary \ref{ldssvec} holds conditioned on $X_{t,1:K}^i$. This can be done efficiently only if $\Delta^*$ can be computed analytically or using some efficient numerical techniques. That will be the focus of future research. But, as discussed earlier, PF-EIS works even if unimodality of $\pss(X_{t,r})$ holds for most particles at most times, i.e. it holds w.h.p. 





We use the temperature tracking problem of Example \ref{temptrack} to explain how to choose $X_{t,s}$. For a given $K$, we would like to choose $X_{t,s}=v_{t,s}$ that makes it most likely for $\pss(v_{t,r})$ to be unimodal.
Given $X_{t-1}^i$, $v_{t,s}^i$, $v_{t,r}$ is a linear function of $C_t$. If $v_{t,r}$ were also a one-to-one function of $C_t$, then one could equivalently find conditions for unimodality of $\pss(C_{t})$, which is easier to analyze. For an approximate analysis, we make it a one-to-one function of $C_t$ by adding a very small variance (compared to that of any $\nu_{t,p}$) noise, $n_{t,s}$, along $B_s$, i.e. given $X_{t-1}^i,v_{t,s}^i$, set $C_t = C_{t-1}^i+B_s v_{t,s}^i + B_r v_{t,r} + B_s n_{t,s}$. Now, $C_t$ is a one-to-one and linear function of $[v_{t,r},n_{t,s}]$. This also makes $\pss(C_{t})$ a non-degenerate pdf.

First consider the case where w.h.p. OL can be multimodal as a function of temperature at only one node $p_0$, for e.g., $h_p(C_{t,p}) = C_{t,p}$, $\forall p \neq p_0$, $\alpha_p^j = 0$, $\forall p \neq p_0$, and either $\alpha_{p_0}^j > 0$ or $h_{p_0}(C_{t,p_0})$ is many-to-one. Then,
\bea
\pss(C_{t}) \se \overbrace{\zeta p(Y_{t,p_0}|C_{t,p_0}) p(C_{t,p_0}|X_{t-1}^i,v_{t,s}^i)}^{\pss(C_{t,p_0})} \times \nn \\
&&  [\prod_{p \neq p_0} p(Y_{t,p}|C_{t,p})] p(C_{t}|C_{t,p_0},X_{t-1}^i,v_{t,s}^i) \ \ \
\label{pssct}
\eea
and the last two terms above are Gaussian (and hence strongly log-concave) as a function of $C_{t,p}, \ p \neq p_0$. If $\pss(C_{t,p_0})$ is also strongly log-concave then $\pss(C_{t})$ (and hence $\pss(v_{t,r})$) will be strongly log-concave, and hence unimodal. Now, $\pss(C_{t,p_0})$ will be strongly log-concave if $\exists \ \eps_0 >0$ such that $\Delta_{C,p_0} = Var[p(C_{t,p_0}|X_{t-1}^i,v_{t,s}^i)] < \inf_{\{ C_{t}: \nabla_{C_{t,p_0}}^2 E_{Y_t}(C_t) < 0 \}} \frac{1}{|\nabla_{C_{t,p_0}}^2 E_{Y_t}(C_t)|+\eps_0}$. This bound can only be computed on the fly. A-priori,  $\pss(C_{t,p_0})$ will be most likely to be log-concave if $v_{t,s}$ is chosen to ensure that $\Delta_{C,p_0}$ is smallest.
 Let $v_{t,s} = v_{t,k_0}$ where the set $k_0$ contains $K$ elements out of $[1,\dots M]$ and $K$ is fixed.  Then, $\Delta_{C,p_0} = \sum_{k \notin k_0} B_{p_0,k}^2 \Delta_{\nu,k}$. This ignores the variance of $n_{t,s}$ (valid since the variance is assumed very small compared to all $\Delta_{\nu,p}$'s). Thus, $\Delta_{C,p_0}$ will be smallest if $v_{t,s}$ is chosen as
\bea
v_{t,s} = v_{t,k_s}, \ k_s \defn \arg \min_{k_0} \sum_{k  \notin k_0} B_{p_0,j}^2 \Delta_{\nu,k}
\label{choosevts}
\eea
When $K=1$, this is equivalent to choosing $k_s = \arg \max_k B_{p_0,k}^2 \Delta_{\nu,k}$. Based on the above discussion, we have the following heuristics.%
\begin{heuristic}
If OL can be multimodal as a function of  temperature at only a single node, $p_0$, and is unimodal as a function of temperature at other nodes, select $v_{t,s}$ using (\ref{choosevts}).
\label{mm1}
\end{heuristic}
\begin{heuristic}
If OL is much more likely to be multimodal as a function of $C_{t,p_0}$, compared to temperature at any other node (e.g. if a sensor at $p_0$ is old so that its failure probability is much larger than the rest), apply Heuristic \ref{mm1} to that $p_0$.
\label{mmmost}
\end{heuristic}
\begin{heuristic}
When $p_0$ is a set (not a single index), Heuristic \ref{mm1} can be extended to select $k_s$ to minimize the spectral radius (maximum eigenvalue) of the matrix, $\sum_{k \notin k_0} B_{p_0,k}B_{p_0,k}^T \Delta_{\nu,k}$.%
\label{mm2}
\end{heuristic}
\begin{heuristic}
If OL is equally likely to be multimodal as a function of any $C_{t,p}$ (e.g. if all sensors have equal failure probability), then $p_0 = [1, \dots M]$. Applying Heuristic \ref{mm2}, one would select the $K$ largest variance directions of STP as $v_{t,s}$.%
\label{equal} 
\end{heuristic}
\begin{heuristic}
If the probability of OL being multimodal is itself very small, then $K=0$ can be used. In Example \ref{temptrack} with all linear sensors, this probability is roughly $1 - \prod_{p,j} (1-\alpha_p^j)$.%
\label{h2}
\end{heuristic}
\begin{heuristic}
For $J=2$ and all linear sensors, $p_0$ may be chosen on-the-fly as $\arg \max_p [(Y_{t,p}^{(1)}-Y_{t,p}^{(2)})^2/\sigma_{obs,p}^2]$ (larger the difference, the more likely it is for OL to be multimodal at that $p$). If the maximum itself is small, set $K=0$.%
\label{onfly}
\end{heuristic}

We show an example now. Consider Example \ref{example1} with $\alpha^{(1)} = \alpha^{(2)} = [0.4,0.01,0.01]$, $\Sigma_\nu = diag([10,5,5])$, $B= [0.95, 0.21, 0.21]'; [-0.21,0.98,-0.05]'; [-0.22, 0, 0.98]'$ (using MATLAB notation). By Heuristic \ref{h2}, the probability of OL being multimodal is about 0.65 which is not small. So we choose $K>0$ ($K=1$). By Heuristic \ref{mmmost}, we choose $p_0=1$ since OL is multimodal as a function of
$C_{t,1}$ with probability 0.64, while that for $C_{t,2}$ or $C_{t,3}$ together is $0.02$ (much smaller).
Applying (\ref{choosevts}) for $p_0=1$, we get $v_{t,s} = v_{t,1}$.

\begin{algorithm*}
\caption{{\small \bf PF-EIS-MT. Going from $\pi_{t-1}^N$ to $\pi_t^N(X_t) = \sum_{i=1}^N \wi \delta(X_{t} - X_{t}^{i})$, \ $X_{t}^{i} = [X_{t,s}^{i},X_{t,r}^{i}]$, \ $X_{t,r}^{i} = [X_{t,r,s}^{i},X_{t,r,r}^{i}]$}}
\label{pfeismt}
\ben
\item
{\em Importance Sample $X_{t,s}$: } $\forall i$, sample $X_{t,s}^{i} \sim p(X_{t,s}^i|X_{t-1}^i)$.
\label{isxts}

\item {\em Efficient Importance Sample $X_{t,r,s}$: } $\forall i$,
\ben
\item Compute  $m_t^i(X_{t-1}^i,X_{t,s}^i,Y_t) =  \arg \min_{X_{t,r}} L^i(X_{t,r})$ and $\Sigma_{IS}^{i}\defn (\nabla^2 L^i(m_t^{i}))^{-1}$ where $L^i$ is defined in (\ref{defL}). Let $m_t^i = \vc{m_{t,s}^i}{m_{t,r}^i}$ and
$\Sigma_{IS}^i = \matr{\Sigma_{IS,s} & \Sigma_{IS,s,r}}{\Sigma_{IS,r} & \Sigma_{IS,r,s}}$.

\item Sample  $X_{t,r,s}^i \sim \n(m_{t,s}^{i}, \ \Sigma_{IS,s}^{i})$.
\een
\label{isxtreismt} 

\item {\em  Mode Track $X_{t,r,r}$: } $\forall i$,
\ben
\item Compute ${m_{t,r}^*}^i$ using (\ref{defsigisrr}). 
\item Set $X_{t,r,r}^i = {m_{t,r}^*}^i$
\een
\label{mtxtreismt} 

\item {\em Weight: }
$\forall i$, compute $w_t^i = \frac{\tilde{w}_t^{i} }{ \sum_{j=1}^N \tilde{w}_t^{j}}$ where  $\tilde{w}_t^{i} = w_{t-1}^{i} \frac{p(Y_t|X_t^{i}) p(X_{t,r}^{i}|X_{t-1}^i, X_{t,s}^i)}{\n(X_{t,r}^{i}; \ m_{t}^{i}, \ \Sigma_{IS}^{i})}$ where $X_{t,r}^i=[X_{t,r,s}^i,X_{t,r,r}^i]$. 
\label{weighteismt}

\item {\em Resample.}  Set $t \leftarrow t+1$ and go to step \ref{isxts}.
\label{weight}
\een
\end{algorithm*} 
\begin{algorithm*}[t!]   
\caption{{\small \bf PF-MT. Going from $\pi_{t-1}^N$ to $\pi_t^N(X_t) = \sum_{i=1}^N \wi \delta(X_{t} - X_{t}^{i})$, \ $X_{t}^{i} = [\tilde{X}_{t,s}^{i},\tilde{X}_{t,r}^{i}]$  }}
\label{pfmt}
\ben
\item
{\em Importance Sample $\tilde{X}_{t,s}$: } $\forall i$, sample $\tilde{X}_{t,s}^{i} \sim p(\tilde{X}_{t,s}^i|X_{t-1}^i)$.
\label{isxtsmt}

\item {\em  Mode Track $\tilde{X}_{t,r}$: } $\forall i$, set $\tilde{X}_{t,r}^i = m_t^{i}$ where $m_t^i(X_{t-1}^i,\tilde{X}_{t,s}^i,Y_t) =  \arg \min_{\tilde{X}_{t,r}} L^i(\tilde{X}_{t,r})$ and $L^i$ is defined in (\ref{defL}).
\label{mtxtrmt}

\item {\em Weight: }  $\forall i$, compute $w_t^i = \frac{\tilde{w}_t^{i} }{ \sum_{j=1}^N \tilde{w}_t^{j}}$ where  $\tilde{w}_t^{i} = w_{t-1}^{i} p(Y_t|X_t^{i}) p(\tilde{X}_{t,r}^{i}|X_{t-1}^i, \tilde{X}_{t,s}^i)$ where  $X_t^i=[\tilde{X}_{t,s}^i,\tilde{X}_{t,r}^i]$.
\label{weightmt}

\item {\em Resample. Set $t \leftarrow t+1$ \& go to step \ref{isxts}.}

\een
\end{algorithm*}

%
%
%

\section{PF-EIS-MT: PF-EIS with Mode Tracker}
\label{pfmtsec}
For any PF (including efficient PFs such as PF-EIS or PF-Doucet), the effective particle size \cite{tutorial,doucet} reduces with increasing dimension, i.e. the $N$ required for a given tracking accuracy increases with dimension. This makes all PFs impractically expensive for LDSS problems. We discuss one possible solution to this problem here.



\subsection{PF-EIS-MT and PF-MT Algorithm}


Consider the LDSS model (\ref{ldssmod}). To apply PF-EIS, we split the state $X_t$ into $[X_{t,s}, X_{t,r}]$, such that $p^*$ is unimodal w.h.p. conditioned on $X_{t,s}$.
As explained earlier, this is ensured if the eigenvalues of $\Sigma_{r}$ are small enough to satisfy (\ref{suff}).
Now, because of the LDSS property, $X_{t,r}$ can further be split into $[X_{t,r,s}, X_{t,r,r}]$ so that the maximum eigenvalue of the covariance of the STP of $X_{t,r,r}$ is small enough to ensure that there is little error in approximating the conditional posterior of $X_{t,r,r}$ by a Dirac delta function at its mode. We call this the Mode Tracking (MT) approximation of importance sampling (IS), or IS-MT. We {\em refer to $\tilde{X}_{t,s} \defn [X_{t,s}, X_{t,r,s}]$ as the ``effective" state and to $\tilde{X}_{t,r} \defn X_{t,r,r}$ as the ``residual" state.} We explain IS-MT in detail below.

In PF-EIS, we IS  $X_{t,s}^i$ from its STP, and we EIS $X_{t,r}^i$ from $\n(m_t^i,\Sigma_{IS}^i)$ where $m_t^i$, $\Sigma_{IS}^i$ are defined in (\ref{defsigis}). Let $m_t^i = \vc{m_{t,s}^i}{m_{t,r}^i}$ and $\Sigma_{IS}^i = \matr{\Sigma_{IS,s} & \Sigma_{IS,s,r}}{\Sigma_{IS,r,s} & \Sigma_{IS,r}}$.
This is equivalent to first sampling $X_{t,r,s}^i \sim \n(m_{t,s}^i, \Sigma_{IS,s}^i)$ and then sampling $X_{t,r,r}^i \sim \n({m_{t,r}^*}^i, \Sigma_{IS,r}^i)$ where
 \bea
{m_{t,r}^*}^i \sdefn m_{t,r}^i + \Sigma_{IS,r,s}^i {\Sigma_{IS,s}^i}^{-1} (X_{t,r,s}^i - m_{t,s}^i), \nn \\
{\Sigma_{IS,r}^*}^i \sdefn \Sigma_{IS,r}^i - \Sigma_{IS,r,s}^i {\Sigma_{IS,s}^i}^{-1} {\Sigma_{IS,r,s}^i}^T
\label{defsigisrr}
\eea
Now, from (\ref{defsigisrr}), ${\Sigma_{IS,r}^*}^i \le \Sigma_{IS,r}^i$.
Also, since $m_t^i$ lies in a locally convex region of $E_{Y_t}(X_{t,s}^i,X_{t,r})$, i.e. $\nabla^2 E_{Y_t} (X_{t,s}^i,m_t^i) \ge 0$ (by Theorem \ref{unimodthm}), $\Sigma_{IS}^i \le \Delta_{r}$. This implies that $\Delta_{r,r}-\Sigma_{IS,r}^i$, which is a square sub-matrix of $\Delta_{r}-\Sigma_{IS}^i$, is also non-negative definite. Thus,
\bea
{\Sigma_{IS,r}^*}^i \le \Sigma_{IS,r}^i  \le \Delta_{r,r}
\label{Deltarrbnd}
\eea
If the maximum eigenvalue of $\Delta_{r,r}$ is small enough, any sample from $\n({m_{t,r}^*}^i, {\Sigma_{IS,r}^*}^i)$ will be close to ${m_{t,r}^*}^i$ w.h.p. So we can set $X_{t,r,r}^i= {m_{t,r}^*}^i$ with little extra error (quantified in the next subsection).
{\em The algorithm is then called PF-EIS-MT. It is summarized in Algorithm \ref{pfeismt}.}
 A more accurate, but also more expensive modification(need to implement it on-the-fly) would be do MT on the low eigenvalue directions of $\Sigma_{IS}^i$. 
%
%
A simpler, but sometimes less accurate, modification is PF-MT (summarized in Algorithm \ref{pfmt}). In PF-MT, we combine $X_{t,r,s}$ with $X_{t,s}$ and importance sample the combined state $\tilde{X}_{t,s} = [X_{t,s},X_{t,r,s}]$ from its STP (or in some cases $X_{t,r,s}$ is empty), while performing mode tracking (MT) on $\tilde{X}_{t,r} = X_{t,r,r}$.

The IS-MT approximation introduces some error in the estimate of $X_{t,r,r}$ (error decreases with decreasing spread of $\pss(X_{t,r,r})$). But it also reduces the sampling dimension from $\dim(X_t)$ to $\dim([X_{t,s} ; X_{t,r,s}])$ (significant reduction for large dimensional problems), thus improving the effective particle size. For carefully chosen dimension of $X_{t,r,r}$, this results in smaller total error, especially when the available number of particles, $N$, is small. This is observed experimentally, but proving it theoretically is an open problem.
We say that the {\em IS-MT approximation is ``valid"} for a given choice of $X_{t,r,r}$ if it results in smaller total error than if it were not used.%

\subsection{IS-MT Approximation}
\label{ismtanal}
We quantify the error due to IS-MT. If we did not use the MT approximation,  $X_{t,r,r}^i \sim \n({m_{t,r}^*}^i, {\Sigma_{IS,r}^*}^i)$.
But using MT, we set $X_{t,r,r}^i = {m_{t,r}^*}^i$. Let the eigenvalue decomposition of ${\Sigma_{IS,r}^*}^i = U {\Lambda_{IS,r}^*}^i U^T$ and let $\lambda_p \defn ({\Lambda_{IS,r}^*}^i)_{p,p}$ be its $p^{th}$  eigenvalue.
Let $d \defn X_{t,r,r}^i - {m_{t,r}^*}^i$. For an $\eps>0$, we bound the probability of $||d|| = ||U^T d|| > \eps$ using Chernoff bounding:%
\bea
Pr(||d|| > \eps) \se Pr(||U^T d|| > \eps) \nn \\
\se Pr( \ e^{s \sum_p (U^T d)_p^2} > e^{s \eps^2} \ ) \nn \\
\sle \prod_p [ \ (1-2 \lambda_p s)^{-1/2}  e^{-s \eps^2/M_{r,r})} \ ] \nn \\
\label{SigISbnd}
\sle [\ (1-2 \lambda_{m} s)^{-1/2}   e^{-s \eps^2/M_{r,r})} \ ]^{M_{r,r}} \ \ \ \ \ \ \\
\label{Sigrrbnd}
\sle [\ (1-2 \Delta_{m} s)^{-1/2}  e^{-s \eps^2/M_{r,r})} \ ]^{M_{r,r}} \ \ \ \ \ \
\eea
where $s > 0$, $\lambda_{m} \defn \max_p \lambda_p$ and $\Delta_m \defn \max_p \Delta_{r,r,p}$.
The first inequality follows by applying Markov inequality, the second follows because $\lambda_p \le \lambda_m, \forall p$ and (\ref{Sigrrbnd}) follows because $\lambda_{m} \le  \Delta_m$ which follows from (\ref{Deltarrbnd}).
Now, (\ref{Sigrrbnd}) holds for any $s>0$ and thus
\bea
&& Pr(||d|| > \eps) \le [ \min_{s>0} \{ (1-2 \Delta_{m} s)^{-1/2}  e^{-s \eps^2/M_{r,r})} ) \} ] ^{M_{r,r}} \nn \\
&& = \left[ ( \frac{M_{r,r} \Delta_m}{\eps^2} )^{-1}  e^{-(\frac{\eps^2}{M_{r,r} \Delta_m} - 1) } \right]^{M_{r,r}/2} \defn B(\Delta_m, \eps)
\eea
Rewriting $[B(\Delta_m, \eps)]^{2/M_{r,r}} = {( \frac{M_{r,r} \Delta_m}{\eps^2} )^{-1} } / { e^{(\frac{\eps^2}{M_{r,r} \Delta_m} - 1)} }$ and applying L'Hospital's rule, we get $\lim_{\Delta_m \tends 0}  B(\Delta_m, \eps) = 0$.
Note that, if instead of (\ref{Sigrrbnd}), we applied $\min_{s>0}$ to (\ref{SigISbnd}), we would get $Pr(||d|| > \eps) \le  B(\lambda_m, \eps)$.
Thus,
\begin{theorem}
Consider any HMM model (satisfying Assumption \ref{hmmass}) and assume that the conditions of Theorem \ref{unimodthm} hold. Let $X_{t,r,r}^i \sim \n({m_{t,r}^*}^i, {\Sigma_{IS,r}^*}^i)$. Then $\lim_{\Delta_m \tends 0} Pr(||X_{t,r,r}^i - {m_{t,r}^*}^i|| > \eps) = 0$ and also $\lim_{\lambda_m \tends 0} Pr(||X_{t,r,r}^i - {m_{t,r}^*}^i|| > \eps) = 0$,
i.e.  $X_{t,r,r}^i$ converges in probability to ${m_{t,r}^*}^i$ in the Euclidean norm as $\Delta_m \defn \max_p \Delta_{r,r,p} \tends 0$ and also as $\lambda_m \defn \max_p ({\Lambda_{IS,r}^*}^i)_{p,p} \tends 0$.
\label{ismtthm}
\end{theorem}
\begin{remark}
Even if the conditions of Theorem \ref{unimodthm} do not hold (inequality (\ref{Deltarrbnd}) does not hold), we can still prove Theorem \ref{ismtthm} if we assume that ${\Sigma_{IS,r}^*}^i = Covar[\pss(X_{t,r,r})]$ (actually $\Sigma_{IS}^i$ is only an approximation to $Covar[\pss(X_{t,r,r})]$). The result will then follow by using the conditional variance identity \cite[Theorem 4.4.7]{stats} to show that $\E_{Y_{t}}[{\Sigma_{IS,r}^*}^i] \le \Delta_{r,r}$.
\end{remark}

In summary, PF-EIS-MT can be used if $\pss(X_{t,r})$ is unimodal w.h.p. and the largest eigenvalue of ${\Sigma_{IS,r}^*}^i$ is small enough to ensure the validity of IS-MT. A sufficient condition is that the largest eigenvalue of $\Delta_{r,r}$ be small enough. The choice of $\eps$ is governed by the tradeoff between the increase in error due to IS-MT and the decrease due to reduced IS dimension. This will be studied in future work.%

%

\subsection{Choosing the MT-Residual states, $X_{t,r,r}$}
\label{chooseKismt}
We first choose an $X_{t,s}, X_{t,r}$ for the EIS step using the unimodality heuristics discussed earlier in Sec. \ref{chooseKunimod}. Then we split $X_{t,r}$ into $X_{t,r,s}$ and $X_{t,r,r}$ so that IS-MT is valid for $X_{t,r,r}$. Then PF-EIS-MT can be implemented with the chosen $X_{t,s}, X_{t,r,s}, X_{t,r,r}$. Alternatively, one can implement PF-MT (faster) with $\tilde{X}_{t,s} =[X_{t,s}; X_{t,r,s}]$, $\tilde{X}_{t,r} = X_{t,r,r}$.
For a given value of $\eps, \eps_2$, two approaches can be used to choose $X_{t,r,r}$. The first is offline and finds the largest $\Delta_m$ so that $B(\Delta_m, \eps) < \eps_2$. The second is online, i.e. at each $t$, for each particle $i$, it finds $\lambda_m$ so that $B(\lambda_m, \eps) < \eps_2$.
\begin{heuristic}
 Begin with $M_{r,r} = M_r$ and keep reducing its value. For each value of $M_{r,r}$, choose the states with the $M_{r,r}$ smallest values of $\Delta_{\nu,r,p}$ (so that $\max_p \Delta_{\nu,r,r,p}$ is smallest) as $X_{t,r,r}$. With this choice, compute $B(\Delta_m, \eps)$ and check if it is smaller than $\eps_2$. If it is smaller, then stop, else reduce $M_{r,r}$ by 1 and repeat the same steps.
A second approach is to do the same thing on-the-fly, using $B(\lambda_m, \eps)$.
\label{hSigrr}
\end{heuristic}

\subsection{Connection with Rao-Blackwellized PF (RB-PF)}
\label{approxrbpf}
%
We first discuss the connection of PF-MT to RB-PF. PF-MT can be interpreted as an approximation of the RB-PF of \cite{gustaffson}. The RBPF of \cite{gustaffson} is applicable when the state vector can be split as $X_t = [X_{t,nl},X_{t,l}]$ with the following property: $X_{t,nl}$ has any general nonlinear or non-Gaussian state space model; but conditioned on $X_{1:t,nl}$, $X_{t,l}$ has a linear Gaussian state space model. Thus the RB-PF of \cite{gustaffson} importance samples $X_{t,nl}$ from its STP but applies the Kalman recursion to compute the conditional prediction and posterior densities (both are Gaussian) of $X_{t,l}$  conditioned on each particle $X_{1:t,nl}^i$. The OL of each particle $X_{1:t,nl}^i$, is computed by marginalizing over the prediction density of $X_{t,l}$.

PF-MT can be understood as an approximation to the RB-PF in the following sense: replace the ``nonlinear" part of the state space by $\tilde{X}_{t,s}$, i.e. $X_{t,nl} \equiv \tilde{X}_{t,s}$, and the ``linear" part by $\tilde{X}_{t,r}$, i.e. $X_{t,l} \equiv \tilde{X}_{t,r}$. In PF-MT, the conditional prediction and posterior densities of $\tilde{X}_{t,r}$ (conditioned on  $\tilde{X}_{1:t,s}^i$) are assumed to be unimodal (not necessarily Gaussian), but narrow. In general, it is not possible to marginalize over any unimodal density.
But if the product of the STP of $\tilde{X}_{t,r}$ and the OL given $\tilde{X}_{t,s}^i$ is narrow enough to be be approximated by its maximum value times a Dirac delta function at its unique maximizer, PF-MT can be interpreted as an RB-PF. In that case, the conditional posterior of $\tilde{X}_{t,r}$ is also approximated by a Dirac delta function. Thus,%
\begin{theorem}
PF-MT (Algorithm \ref{pfmt}) is  RB-PF (Algorithm 1 of \cite{gustaffson}) with the following approximation at each $t$:
\bea
&& p(Y_t|\tilde{X}_{t,s}^i, \tilde{X}_{t,r}) p (\tilde{X}_{t,r}|X_{t-1}^i, \tilde{X}_{t,s}^i)   \nn \\
\label{pfmtapprox}
& & = p(Y_t|\tilde{X}_{t,s}^i, \tilde{X}_{t,r}^i) p (\tilde{X}_{t,r}^i|X_{t-1}^i, \tilde{X}_{t,s}^i) \delta(\tilde{X}_{t,r}-\tilde{X}_{t,r}^i) \\ 
&& \tilde{X}_{t,r}^i = m_t^i = \arg \max_{\tilde{X}_{t,r}} [ p(Y_t|\tilde{X}_{t,s}^i, \tilde{X}_{t,r}) p (\tilde{X}_{t,r}|X_{t-1}^i, \tilde{X}_{t,s}^i) ] \nn
\eea
With the above approximation, the following also holds:
\bea
\pss(\tilde{X}_{t,r}) \defn p(\tilde{X}_{t,r}|X_{t-1}^i,\tilde{X}_{t,s}^i,Y_t) = \delta(\tilde{X}_{t,r} - m_t^i) \ \ \
\eea
\vspace{-0.2in}
\label{pfmtrbpf}
\end{theorem}
 The proof is a simple exercise of simplifying RB-PF expressions using (\ref{pfmtapprox}) and hence is omitted.

For PF-EIS-MT, replace $\tilde{X}_{t,r}$ by $X_{t,r,r}$ and $\tilde{X}_{t,s}$ by $[X_{t,s};X_{t,r,s}]$ in the above discussion. Also, importance sampling from the STP in case of RB-PF is replaced by EIS.
%
%

\section{Relation to Existing Work}
\label{relwork}
We discuss here the relation of our algorithms to existing work.
The problem of estimating temperature at a large number of locations in a room using a network of sensors is also studied in \cite{mouraRC, moura_network}. Their focus is on modeling the spatio-temporal temperature variation using an RC circuit, estimating its parameters, and using the model for predicting temperature at unknown nodes. They assume zero sensor failure probability and observation noise (usually valid when sensors are new) and hence do not require tracking. In a practical system, one can use \cite{mouraRC} when sensors are new and reliable, but track the temperature using PF-EIS-MT (and the models estimated using \cite{mouraRC}) when sensors grow older and unreliable.

\begin{figure*}[t!]
\subfigure[Sensor failure (temperature independent)] 
{\label{pfeisnew2N100}
\begin{tabular}{c}
\epsfig{file = 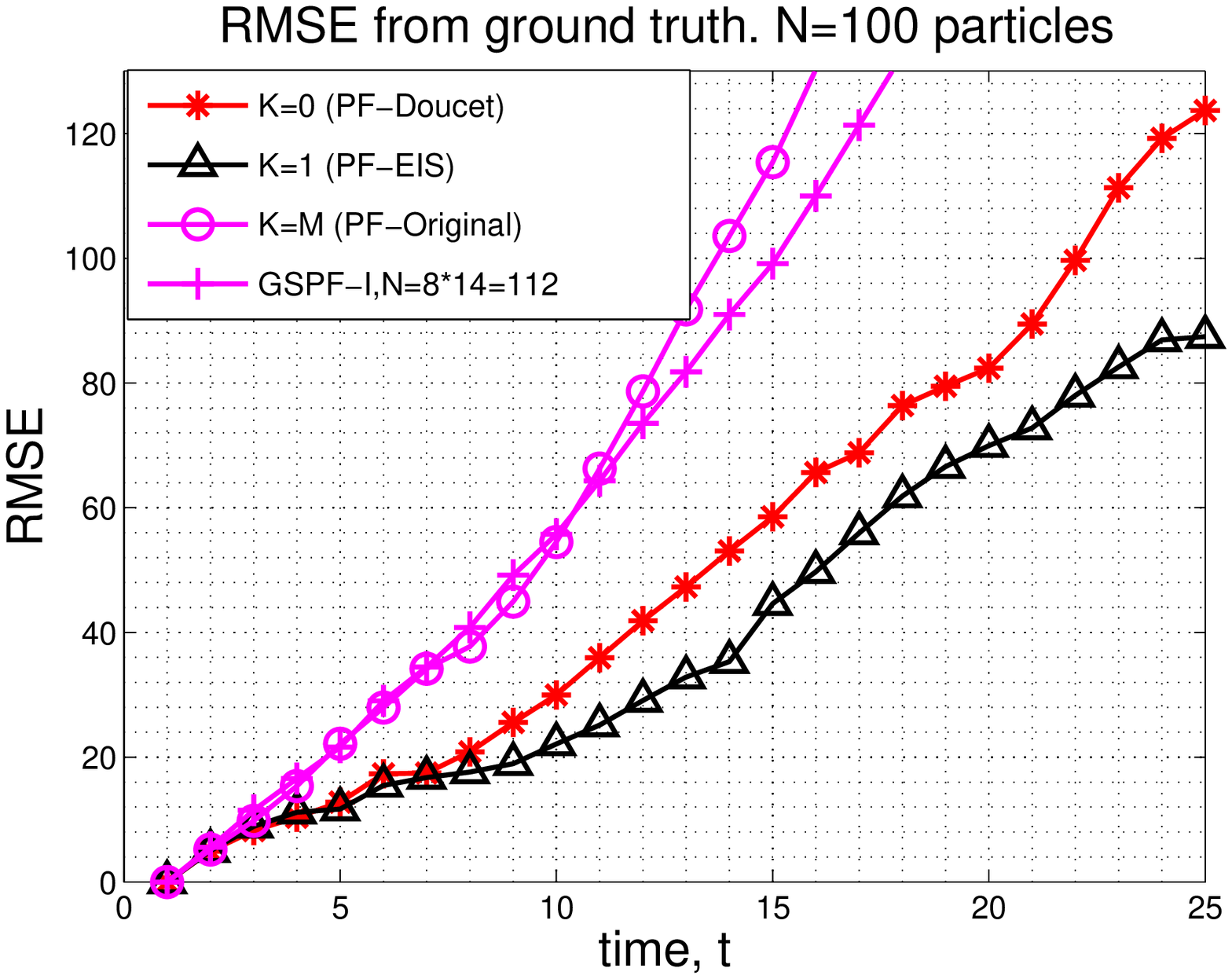, width=5.5cm,height=4cm} \\
\epsfig{file = 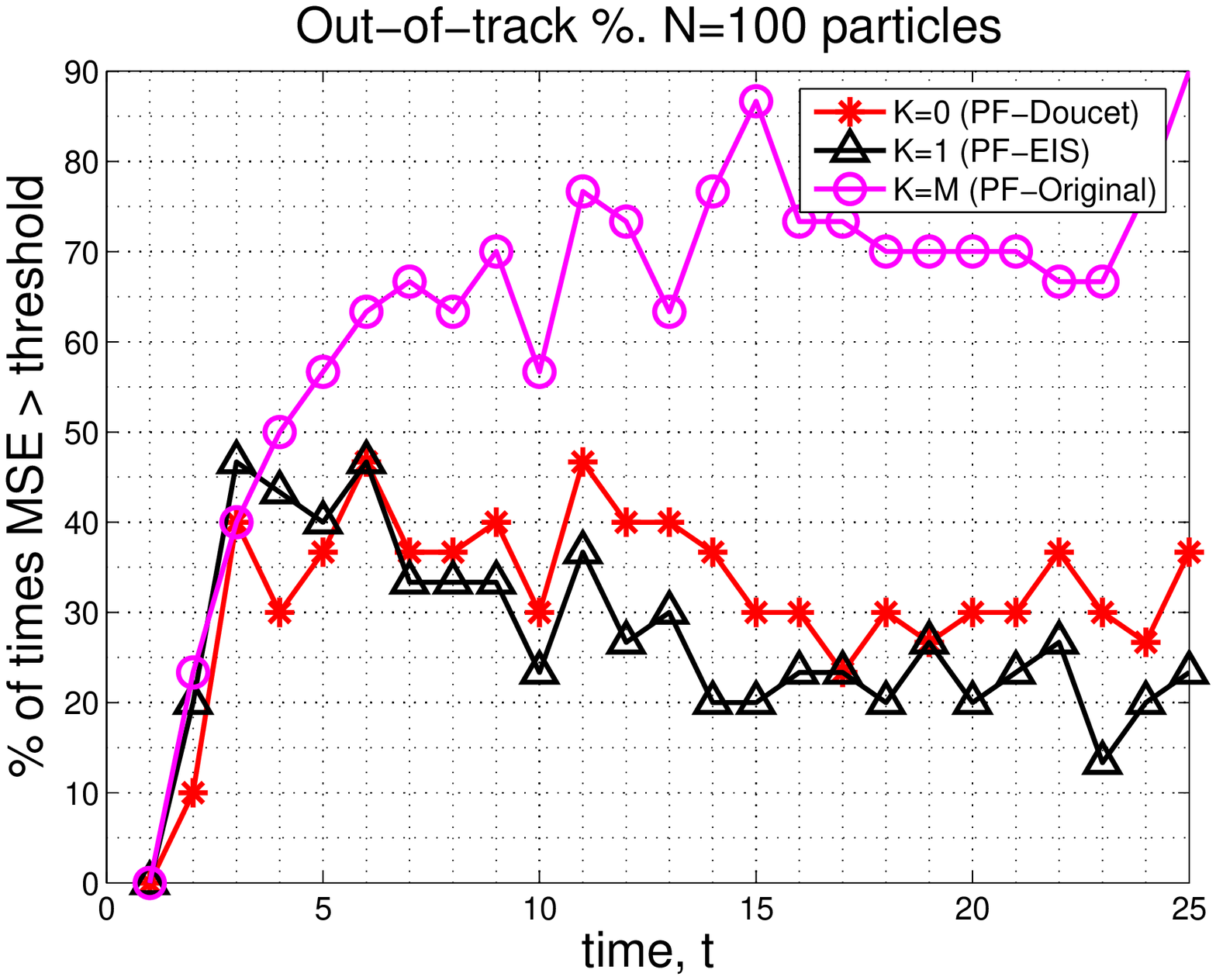, width=5.5cm,height=4cm}
\end{tabular}
}
\subfigure[Sensor failure (weakly temperature dependent)]
{\label{outlier_all}
\begin{tabular}{c}
\epsfig{file = 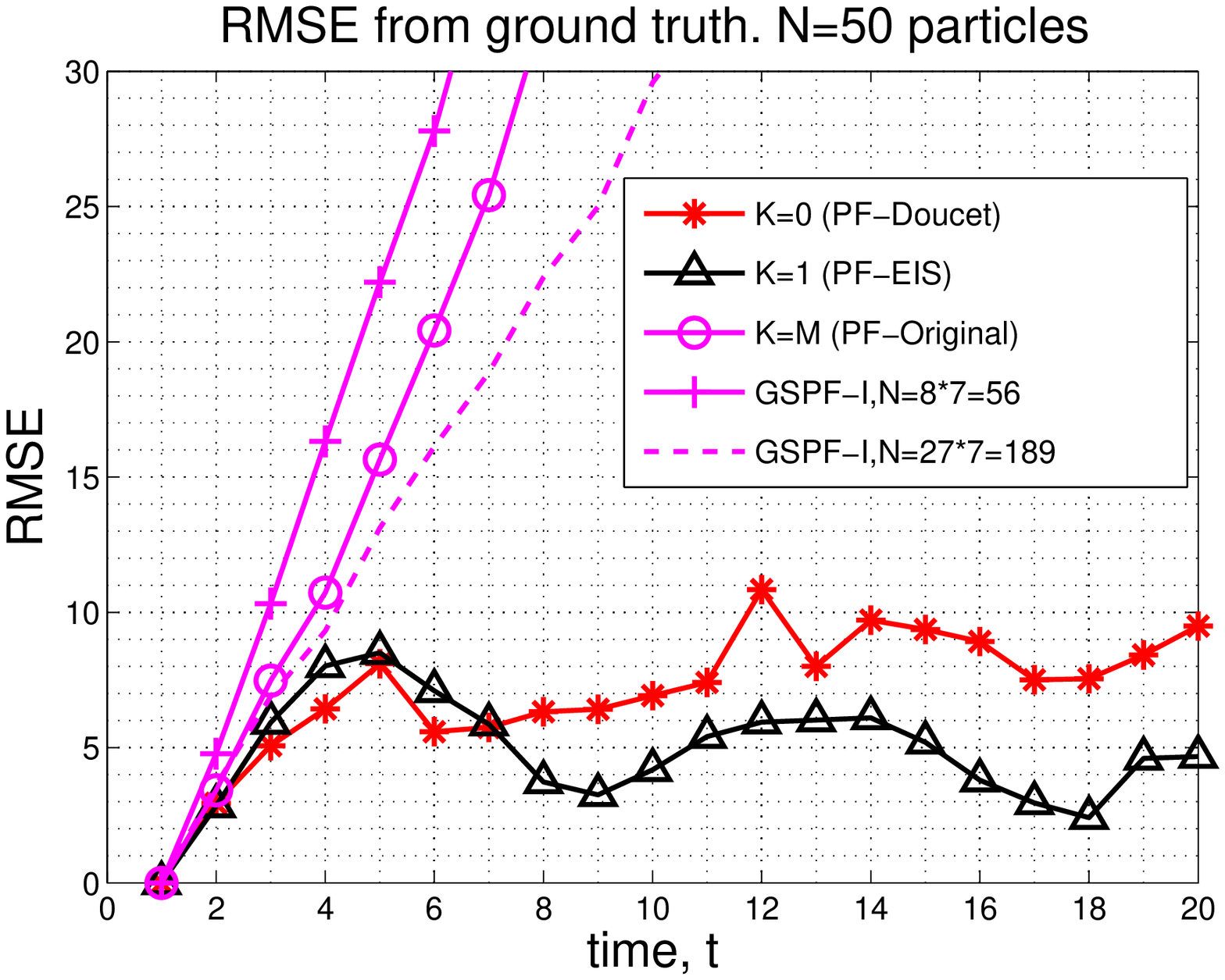, width=5.5cm,height=4cm} \\
\epsfig{file = 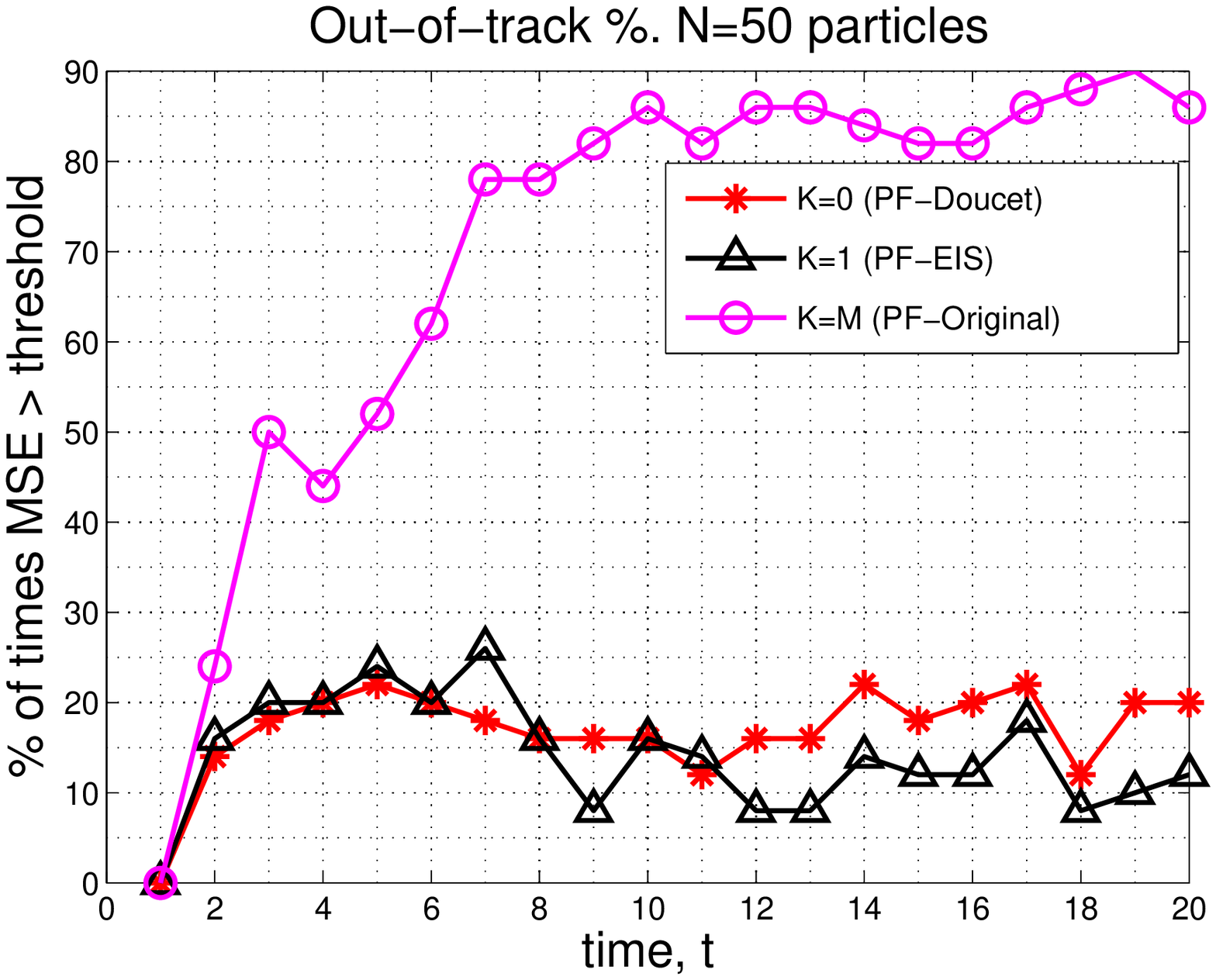, width=5.5cm,height=4cm}
\end{tabular}
}
\subfigure[Squared sensor at node 1] 
{\label{Ctsqr}
\begin{tabular}{c}
\epsfig{file = 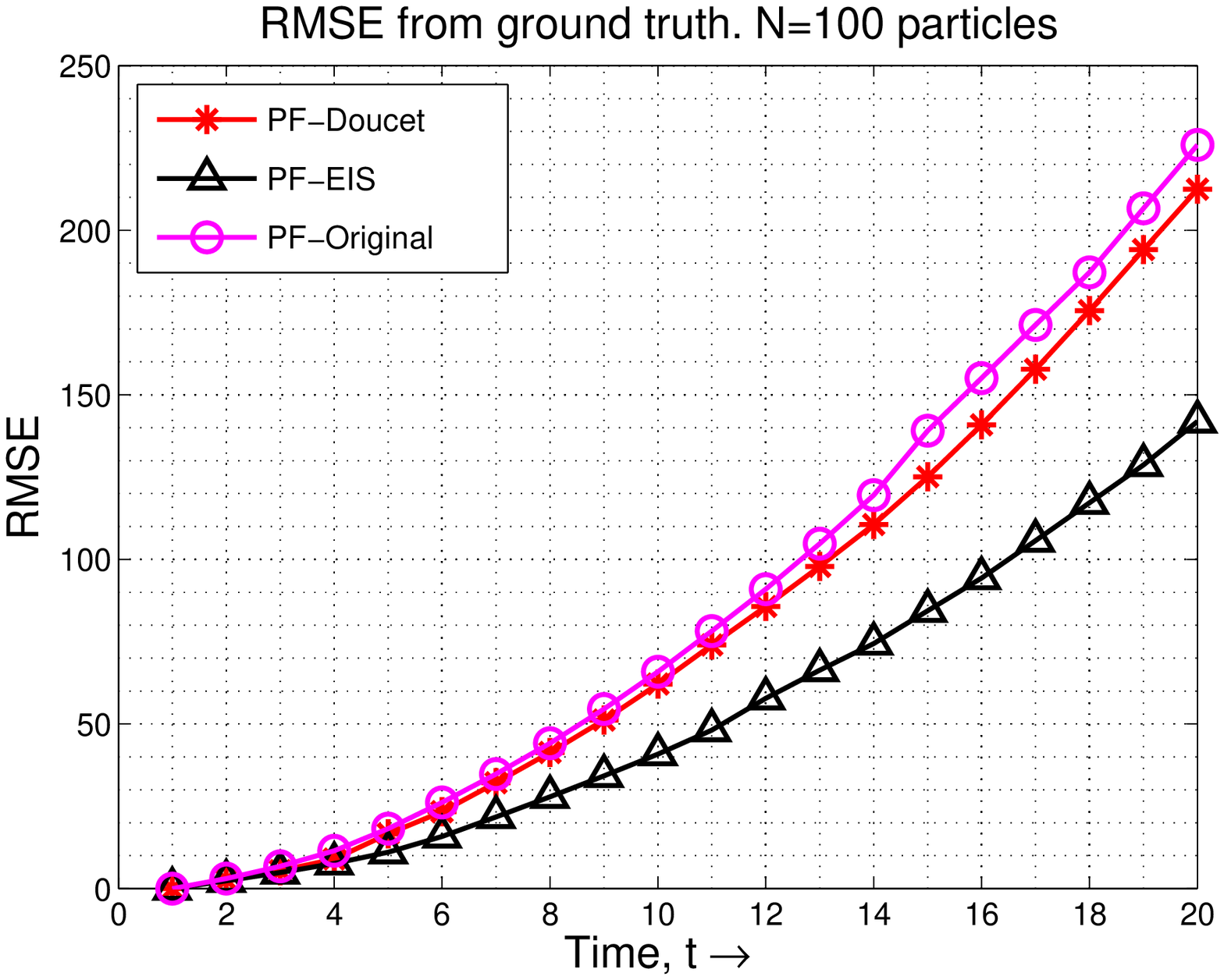, width=5.5cm,height=4cm} \\
\epsfig{file = 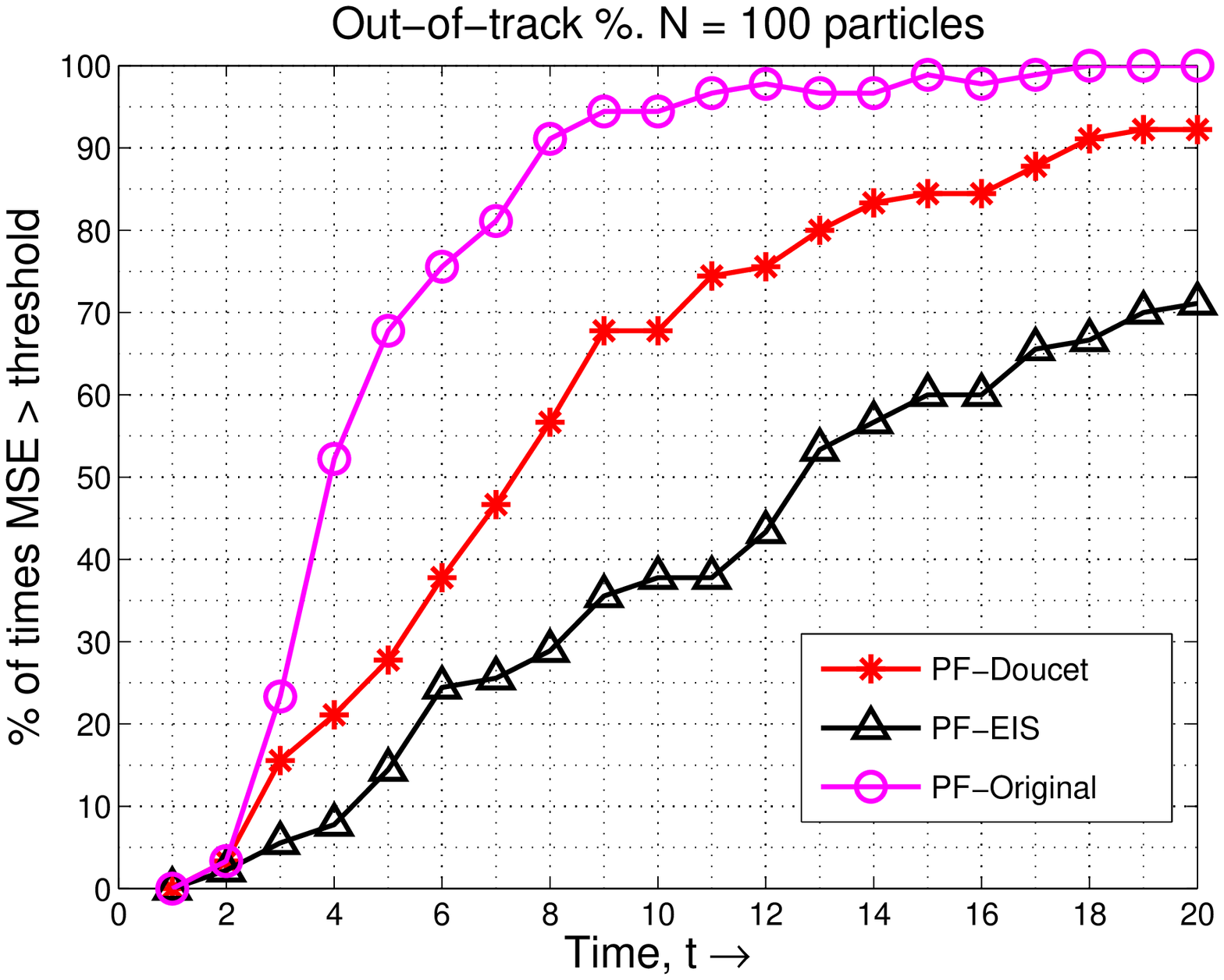, width=5.5cm,height=4cm}
\end{tabular}
}
\subfigure[Table of parameters]
{\small
\centerline{
\begin{tabular}{|l|l|l|l|l|l|l|l|l|l|}
\hline
Fig. & $h(C_t)$ & $p_f(Y_{t,p}^{(j)} | C_{t,p})$ & $\alpha^{(1)}, \alpha^{(2)}$ & $\sigma_{obs}^2$ & $\Delta_\nu$ & $B$ & $C_0$ & $a$ & N \\
\hline
\ref{pfeisnew2N100} & $h(C_t)=C_t$ & $\n(0,100)$ & $\alpha^{(1)} = [.9, .1, .1]$, & [10, 1, 1] & $diag(10,5,5)$, & $[.99, .1, .1]';$ &  $[0, 0, 0]'$ & $1$ & 100 \\
& & & $\alpha^{(2)} = [.4, .01, .01];$ &  &  & $[-.10, 0.99, -.01]';$  & & & \\
 &  & & & & &  $[-.10, 0,.99]'$ & & & \\
\hline
\ref{outlier_all} &  $h(C_t)=C_t$   & $\n(.2C_{t,p},100)$ & $\alpha^{(1)} =  [.4, .01, .01]$ & $[1,1,1]$ & $diag(10,5,5)$ & $[.95, .21, .21]';$ & $[0, 0, 0]'$ & $1$ & 50 \\
 &  & & $\alpha^{(2)} =  [.4, .01, .01]$ & &   &  $[-.21,.98,-.05]';$ & & & \\
 &  & & & & &  $[-.22, 0, .98]'$ & & & \\
\hline
\ref{Ctsqr} & $h_1(C_t)= C_{t,1}^2$   &  & $\alpha^{(1)} = [0, 0, 0]$ & $[3, 1, 1]$ & $diag(10,5,5)$, & $[.95, .21, .21]';$ & $[5, 5, 5]'$ & $.7$ & 50 \\
& $h_p(C_t)= C_{t,p}^2$, $p>1$ & & & &   &  $[-.21,.98,-.05]';$  & & &  \\
 &  & & & & &  $[-.22, 0, .98]'$ & & & \\
\hline
\end{tabular}
\label{fig1parameters}
}
}
\caption{{\small
Comparing RMSE, out-of-track \% and $N_{eff}$ of PF-EIS (black-$\triangle$) with that of  PF-Doucet (red-*), PF-Orig (magenta-o) and GSPF-I (magenta -+). 
RMSE at time $t$ is the square root of the mean of the squared error between the true $C_t$ and the tracked one ($N$-particle PF estimate of $\E[C_t|Y_{1:t}]$).
Out-of-track \% is the percentage of realizations for which the norm of the squared error exceeds an in-track threshold (2-4 times of total observation noise variance). In-track threshold for Fig. \ref{pfeisnew2N100} was 48, for Fig. \ref{Ctsqr} was 20 and for Fig. \ref{outlier_all} was 12. We averaged over 90 Monte Carlo simulations in Figs. \ref{outlier_all} and \ref{Ctsqr} and over 40 in Fig. \ref{pfeisnew2N100}.
Note $C_0$ refers to the starting value of $C_t$.
}}
\label{pfeisfigs}
\end{figure*}

For multimodal OL or STP, if there are only a few modes at known mode locations, the Gaussian Sum PFs (GSPF-I or GSPF-II) of \cite{gspf} can be used.
All examples shown in \cite{gspf} have a one dimensional process noise, and thus effectively a one dimensional state. As dimension increases, the number of mixands that need to be maintained by GSPF-I increases significantly.  We compare PF-EIS with GSPF-I in Fig. \ref{pfeisfigs}. GSPF-II defines a mixand about each possible mode of OL or of STP, followed by resampling to prune insignificant modes. The possible number of OL modes increases with dimension, even though for a given observation, it is highly unlikely that all modes appear. For e.g., in case of  tracking temperature at 50 nodes with 2 sensors per node, each with nonzero failure probability, the maximum number of {\em possible} OL modes at any time is $2^{50}$. Another work that also approximates a multimodal pdf by a mixture density is \cite{berger93}.
The Independent Partition PF (IPPF) of \cite{orton} and the IPPF-JMPD of \cite{kreucher} propose efficient PFs for multiple target tracking. There the motion model of different targets is independent, while the OL is coupled when the targets are nearby (because of correspondence ambiguity between observations and targets). The main idea of IPPF is to resample independently for each target when the targets are significantly far apart (their OLs are roughly independent). 
In our work, and also in other LDSS problems, this cannot be done since the temperature (or other state) dynamics of different nodes is coupled (temperature change is spatially correlated). 

The main idea of MT was first introduced by us in \cite{pami07} and first generalized in \cite{pap1,pap2,icassp07}.
The work of \cite{turbopf} which proposes a ``PF using gradient proposal" is related to \cite{pami07}. 
The MT step can also be understood as Rao-Blackwellization \cite{chen_liu,gustaffson} if the approximation of Theorem \ref{pfmtrbpf} holds.
Another recent PF that also performs approximate marginalization, but only on the stable directions of the state space, is \cite{chorin}. This can be made more efficient by using the EIS idea on the unstable directions. Many existing algorithms may be interpreted as special cases of PF-EIS-MT, for e.g.  PF-Original is PF-EIS-MT with $X_{t,s}=X_t$, PF-Doucet is PF-EIS-MT with $X_{t,r,s}=X_t$, and the approximate ``posterior mode tracker"  of \cite{YezziSoatto2004} is approximately PF-EIS-MT with $X_{t,r,r}=X_t$.
%

There is a fundamental difference between MT and the commonly used idea of replacing the PF estimate of the posterior by a Dirac delta function at the highest weight particle (or at the mode of the PF posterior estimate), as in \cite{wilsky}, or doing this for a subset of states, as in \cite{kevin}. The latter can be understood as an extreme type of resampling which will automatically occur in any PF if the largest weight particle has much higher weight than any other particle. It still requires IS on the entire state space to first get the PF estimate of posterior.

\begin{figure*}[t!]
\centerline{
\subfigure[Sensor failure (temperature independent)]
{\label{pfmt3_N100fig}
\begin{tabular}{cc}
\epsfig{file = 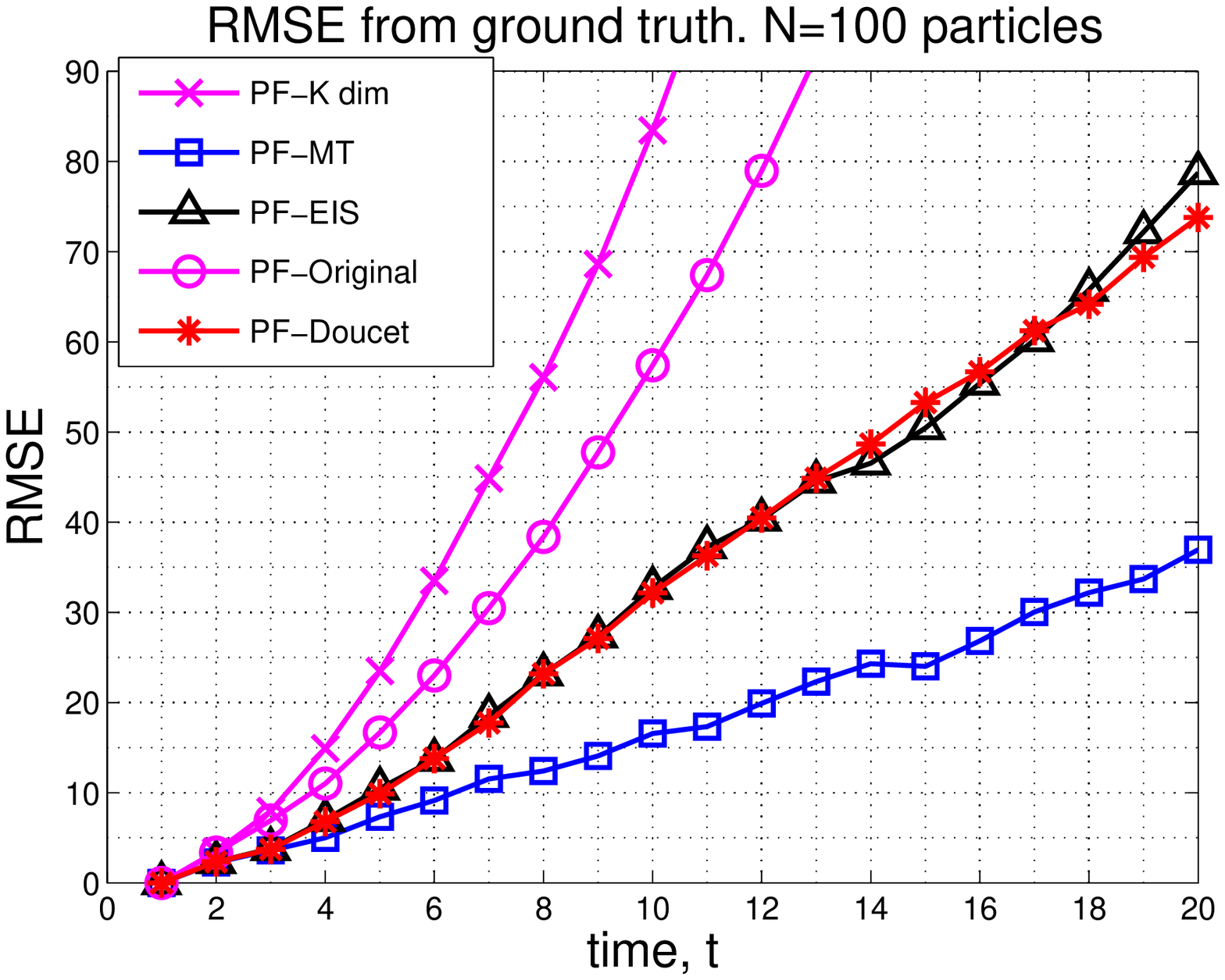,  width=5.5cm,height=4.0cm} &
\epsfig{file = 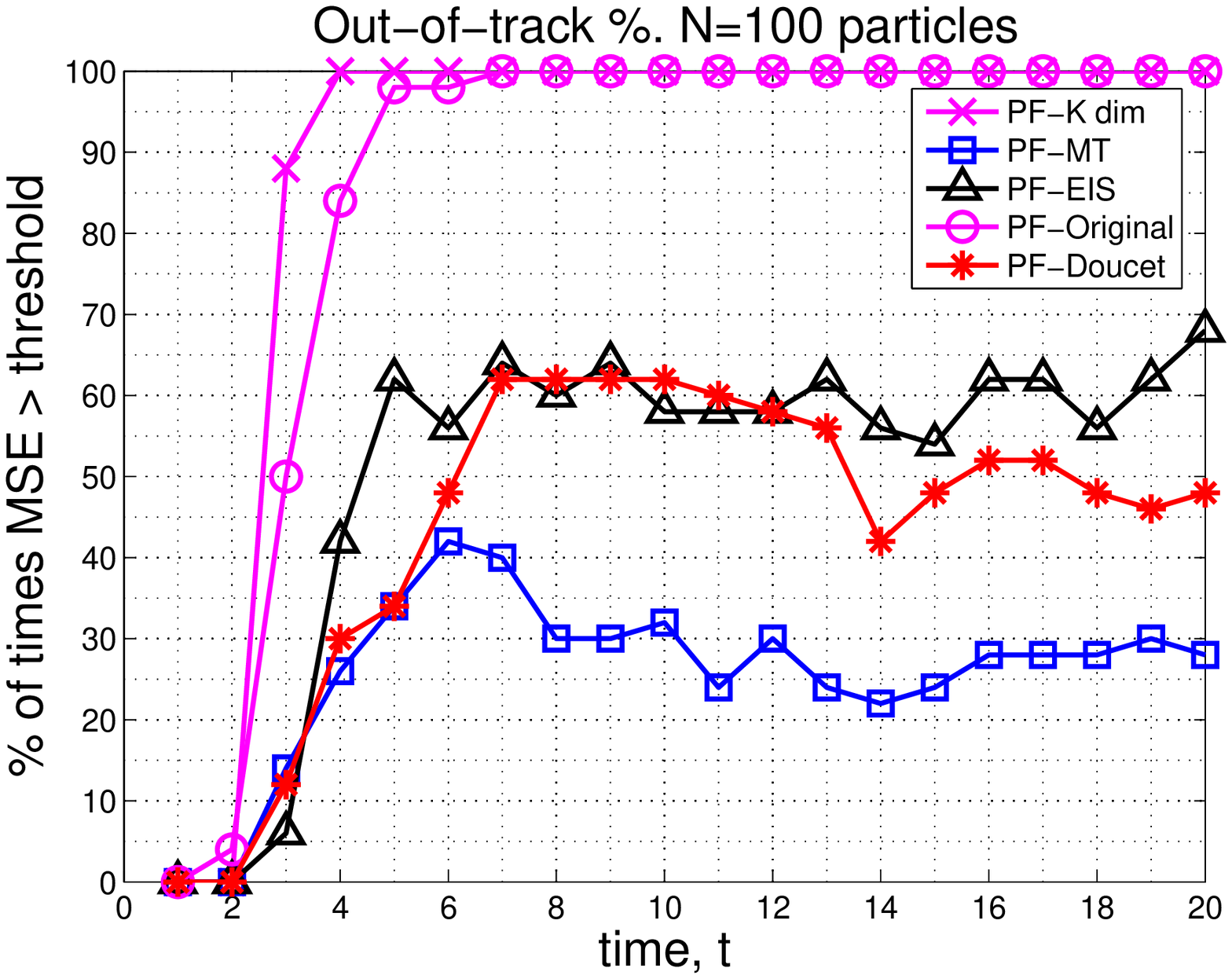,  width=5.5cm,height=4.0cm}
\end{tabular}
}
\subfigure[Robustness to model error] 
{\label{pfmtrobust}
\begin{tabular}{c}
\epsfig{file = 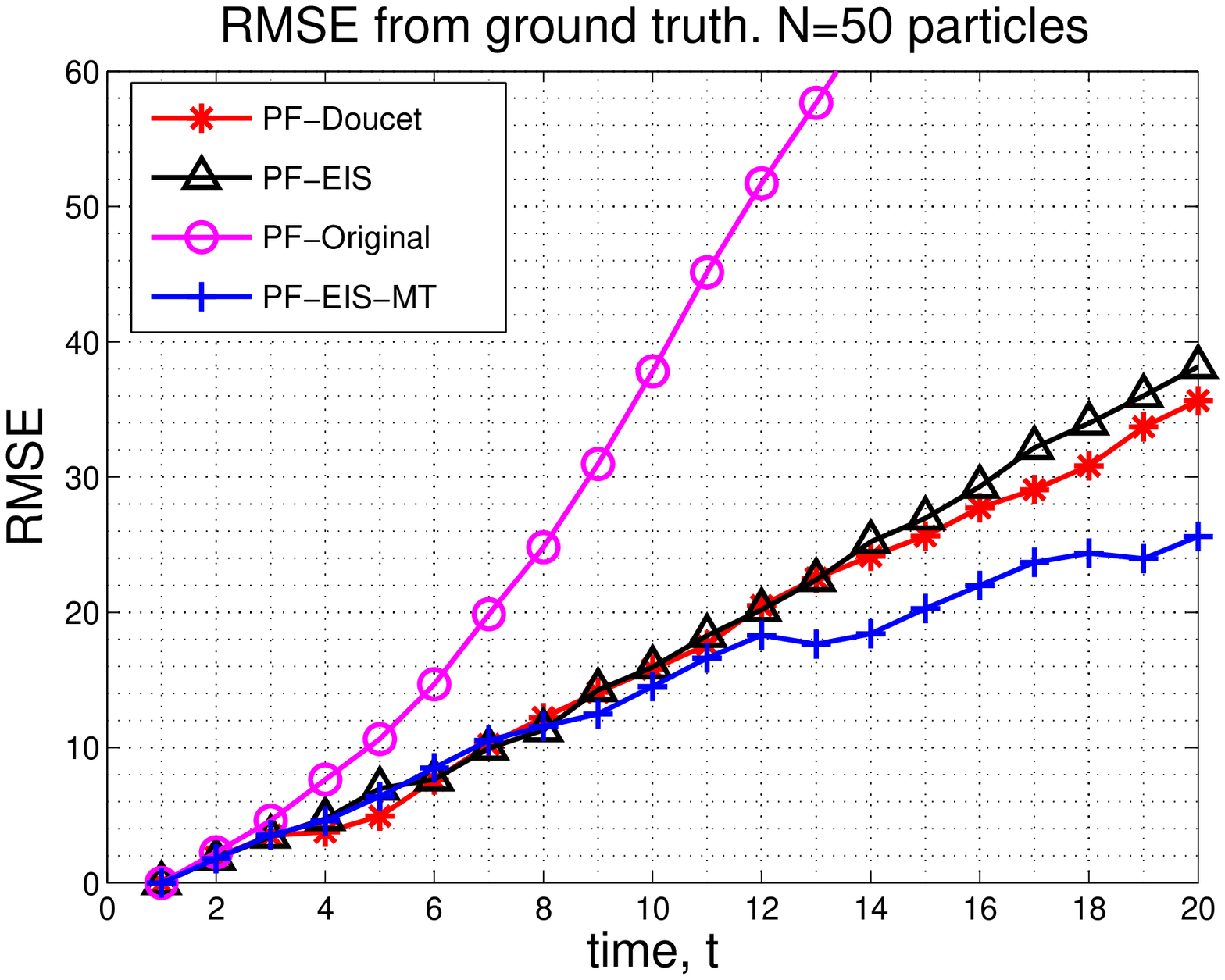, width=5.5cm,height=4.0cm}
\end{tabular}
}
}
\subfigure[Table of parameters. The notation $b_k$ denotes a row vector of $bs$ of length $k$, e.g. $1_9$.]
{\small
\centerline{
\begin{tabular}{|l|l|l|l|l|l|l|l|l|l|l|}
\hline
Fig. & $M$ & $h(C_t)$ & $p_f(Y_{t,p}^{(j)} | C_{t,p})$ & $\alpha^{(1)}, \alpha^{(2)}$ & $\sigma_{obs}^2$ & $\Delta_\nu$ & $B_{1,:}$ & $C_0$ & $a$ & $N$ \\
\hline
\ref{pfmt3_N100fig}& $10$ &  $h(C_t)=C_t$   & $\n(0,100)$ & $\alpha^{(1)} = [0.9, 0.01_9]$ & $1_{10}$ & $diag([10,1_9])$ & $[0.83, 0.18_9 ]'$ & $[0_{10}]'$ & $1$ & 100 \\
& &  & & $\alpha^{(2)} = [0.4, 0.01_9]$ & &   & & & &  \\
\hline
 \ref{pfmtrobust}& $5$ &  $h(C_t)=C_t$   & $\n(0.2C_{t,p},100)$ & $\alpha^{(1)} =\alpha^{(2)} = [0.2, 0.1_4]$ & $[1_{5}]$ & $diag([5,5,1_3])$ & $[0.7, 0.35_5 ]'$ ; & $[0_5]'$ & $1$ & 50 \\
 &  &  & & $\alpha^{(1)}_{sim}=[0.95, 0.1_4]$ & & &  &  & &  \\
\hline
\end{tabular}
}
\label{fig3parameters}
}
\caption{\small{Comparing PF-MT (blue-$\square$) in \ref{pfmt3_N100fig} and PF-EIS-MT (blue-+) in \ref{pfmtrobust} with PF-Doucet (red-*), PF-EIS (black-$\triangle$), PF-Orig (magenta-o) and PF-Orig-K dim (magenta-x).
In Fig. \ref{pfmt3_N100fig},  $M=10$ was used.  $X_{t,s}=v_{t,1}$ was used for both PF-EIS and PF-MT. Averaged over 50 simulations. PF-MT has best performance. In Fig. \ref{pfmtrobust}, we test the robustness to error in the failure probability parameter. $M=5$ was used. We used $X_{t,s}=v_{t,1}$, $X_{t,r,s}=v_{t,2}$ for PF-EIS-MT. $X_{t,s}=v_{t,1}$ was used for PF-EIS. Averaged over 100 simulations.
PF-EIS-MT is the most robust when $N=50$ particles were used (available $N$ is small). If $N=100$ particles are used, PF-EIS is the most robust (not shown). %
}}
\label{pfmtfigs}
\end{figure*}

\section{Simulation Results}
\label{sims}
We used Root Mean Squared Error (RMSE) of the PF approximation of the MMSE state estimate (from its true value) and percentage of out-of-track realizations to compare the performance of PF-EIS with that of PF-Original (PF-EIS with $K=M$) \cite{gordon} and  PF-Doucet (PF-EIS with $K=0$) \cite{doucet} in Fig. \ref{pfeisfigs}. The number of particles ($N$) was kept fixed for all PFs in a given comparison. We also show the RMSE plot of GSPF-I \cite{gspf} with total number of particles (number of mixtures times number of particles per mixture) roughly equal to $N$.
In Fig. \ref{pfmtfigs}, we show superior performance of PF-MT and PF-EIS-MT over PF-EIS, PF-Doucet, PF-Original and PF-Orig-$K$-dim (dimension reduced original PF, i.e. original PF run on only the first $K$ dimensions). 

Note that for multimodal posteriors, the RMSE at the current time does not tell us if all significant modes have been tracked or not. But, if a significant mode is missed, it will often result in larger errors in future state estimates, i.e. the error due to the missed mode will be captured in future RMSEs. In many problems, the goal of tracking is only to get an MMSE state estimate, and not necessarily view all the modes, and in these cases RMSE is still the correct performance measure. If a missed posterior mode does not result in larger future RMSEs, it does not affect performance in any way\footnote{The true posterior is unknown. The only other way to evaluate if a PF is tracking all the modes at all times, is to run another PF with a very large number of particles and use its posterior estimate as the true one.}. 
 Of course, the increase in error due to a missed mode may occur at different time instants for different realizations and hence the average may not always truly reflect the loss in tracking performance.

{\em Evaluating  PF-EIS: }
We first explain a typical situation where PF-Doucet fails but PF-EIS does not. This occurs when the STP is broad and the OL is bimodal (or in general, multimodal) with modes that lie close to each other initially, but slowly drift apart. PF-Doucet uses gradient descent starting at $C_{t-1}^i$ to find the mode. When $p^*$ is multimodal, it approximates $p^*$ by a Gaussian about the mode in whose basin-of-attraction the previous particle (i.e. $C_{t-1}^i$) lies. At $t=0$, particles of $v_t$ are generated from the initial state distribution and so there are some particles in the basin-of-attraction of both modes. But due to resampling, within a few time instants, often all particles cluster around one mode. If this happens to be the wrong mode, it results in loss of track. In contrast, PF-EIS samples  $v_{t,s}$ from its STP, i.e. it generates new particles near both OL modes at each $t$, and so does not lose track.

All plots of Fig. \ref{pfeisfigs} simulated Example \ref{temptrack} with $M=3$. Model parameters used for each subfigure are given in the table in Fig. \ref{fig1parameters}. The example of Fig. \ref{pfeisnew2N100} is a special case of Example \ref{example1}. It has $M=3$ sensor nodes; $J=2$ sensors per node; all linear sensors and ``temperature-independent failure", i.e. $p_f(Y_{t,p}^{(j)}|C_{t,p}) = p_f(Y_{t,p}^{(j)}) = \n(Y_{t,p}^{(j)}; 0, 100)$. Temperature change followed a random walk model, i.e. $a=1$. By Heuristic \ref{mmmost}, we choose $p_0=1$ since OL is multimodal as a function of $C_{t,1}$ with much higher probability than at other nodes (we simulate an extreme case). Applying (\ref{choosevts}) for $p_0=1$, we get $v_{t,s} = v_{t,1}$. This was used for PF-EIS.
As can be seen, RMSE for PF-EIS was smaller than for PF-Doucet and so were the number of ``out of track" realizations. GSPF-I \cite{gspf} with $G=8$ mixtures and $Ng=7$ particles per mixture (a total of 56 particles) and PF-Original had much worse performance for reasons explained earlier (used inefficient importance densities).

In Fig. \ref{outlier_all}, we simulated ``weakly temperature dependent sensor failure", i.e. $p_f(Y_{t,p}^{(j)}|C_{t,p}) = \n(Y_{t,p}^{(j)}; 0.2 C_{t,p}, 100 \sigma_{obs,p}^2)$. Also, sensor failure probability at node 1 was lower than in Fig. \ref{pfeisnew2N100}. Thus the performance of all algorithms is better. 

Fig. \ref{Ctsqr} used $J=1$ sensor per node and a squared sensor at node 1, i.e. $h(C_t) = [C_{t,1}^2; C_{t,2} ; C_{t,3}]$. All sensors had zero failure probability, i.e. $\alpha_p^{(1)} = 0, \forall p$.  Temperature change followed a first order autoregressive model\footnote{This example is a difficult one because OL is almost always bimodal with two equal modes. With a random walk model on $v_t$, even $N=100$ particles were not enough for accurate tracking using any PF.} with $a=0.7$. In this case OL is bimodal as a function of $C_{t,1}$ whenever $Y_{t,1}$ is significantly positive. This happens w.h.p when temperatures are greater than $\sqrt{3\sigma_{obs,1}}= 2.3$ (or less than $-2.3$) which itself happens very often. Also, often, the modes are initially nearby and slowly drift apart as the magnitude of $Y_{t,1}$ increases. As explained earlier, this is just the situation that results in failure of PF-Doucet. Performance of PF-Doucet is significantly worse than that of PF-EIS (which used $v_{t,s} = v_{t,1}$ obtained by applying (\ref{choosevts}) for $p_0=1$). Note that we initiated tracking with an initial known temperature of $5$, so that there was a bias towards positive temperature values and it was indeed possible to correctly track the temperature and its sign.

\begin{figure*}[t!]
\centerline{
\subfigure[$N_{eff}$ for Fig. \ref{pfeisnew2N100}]
{\label{Neff1}
\begin{tabular}{c}
\epsfig{file = 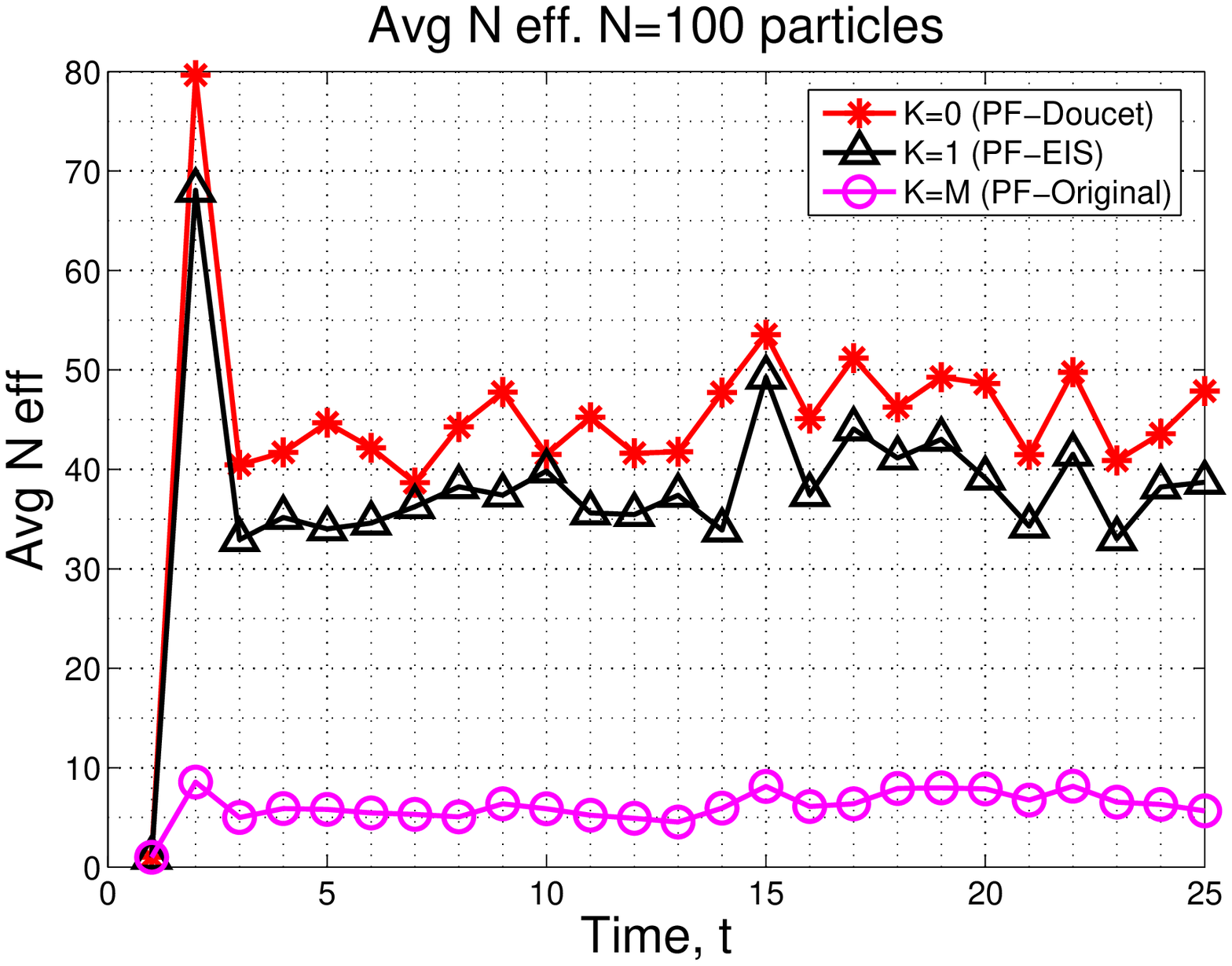, width=4.25cm,height=4cm}
\end{tabular}
}
\subfigure[$N_{eff}$ for Fig. \ref{outlier_all}]
{\label{Neff2}
\hspace{-0.28in}
\begin{tabular}{c}
\epsfig{file = 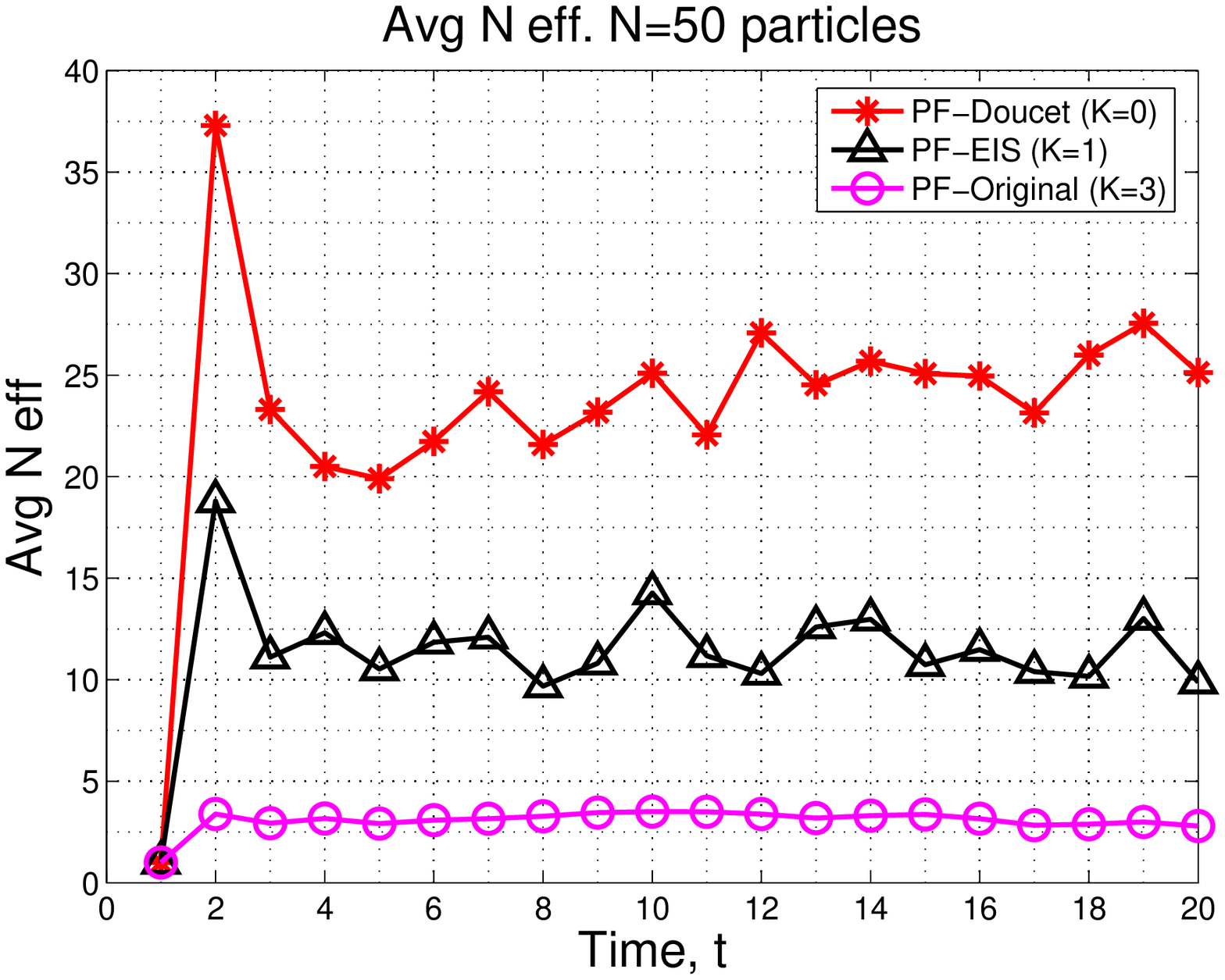, width=4.25cm,height=4cm}
\end{tabular}
}
\hspace{-0.28in}
\subfigure[$N_{eff}$ for Fig. \ref{Ctsqr}]
{\label{Neff3}
\begin{tabular}{c}
\epsfig{file = 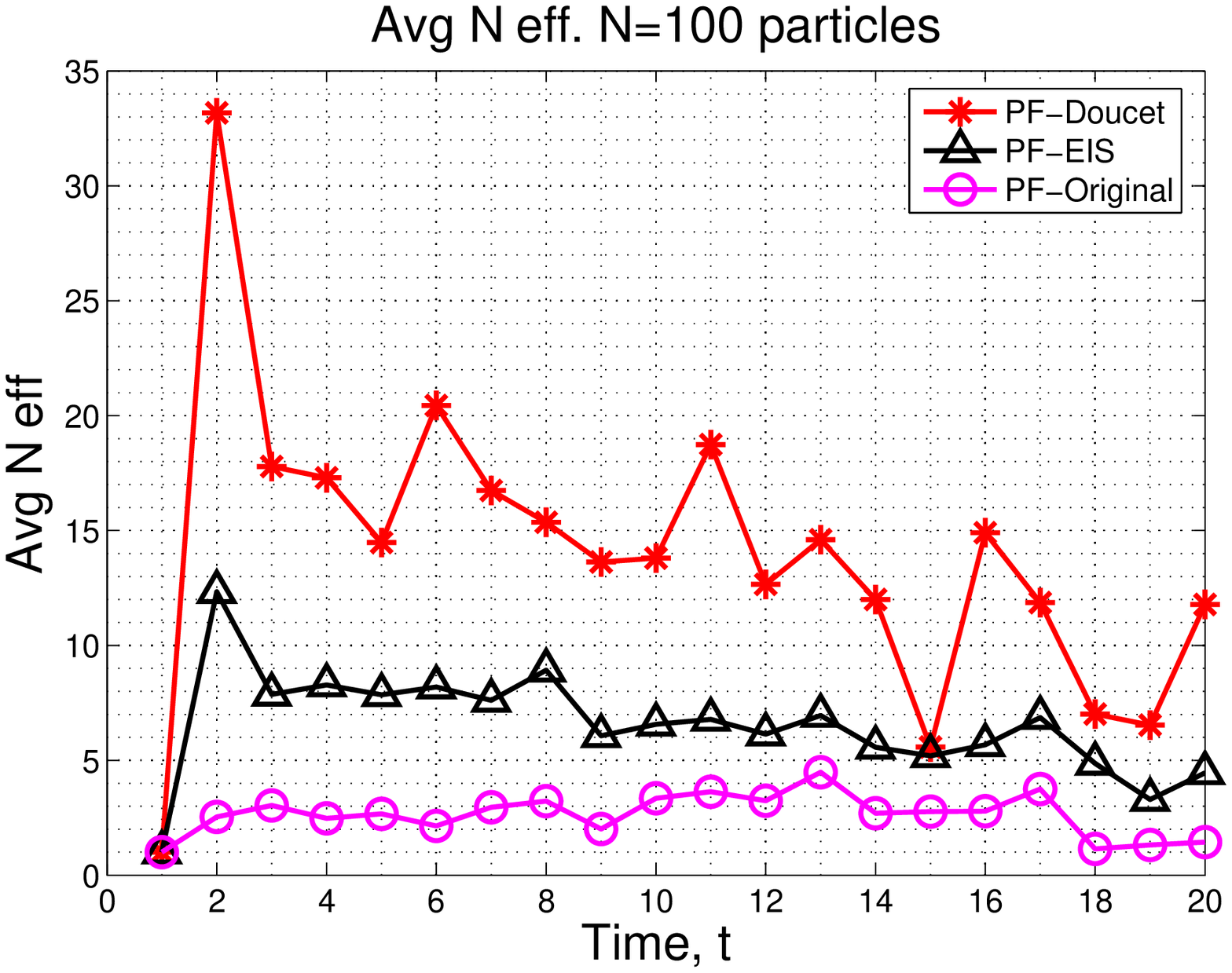, width=4.25cm,height=4cm}
\end{tabular}
}
\hspace{-0.28in}
\subfigure[$N_{eff}$ for Fig. \ref{pfmt3_N100fig}]
{\label{Neff4}
\begin{tabular}{c}
\epsfig{file = 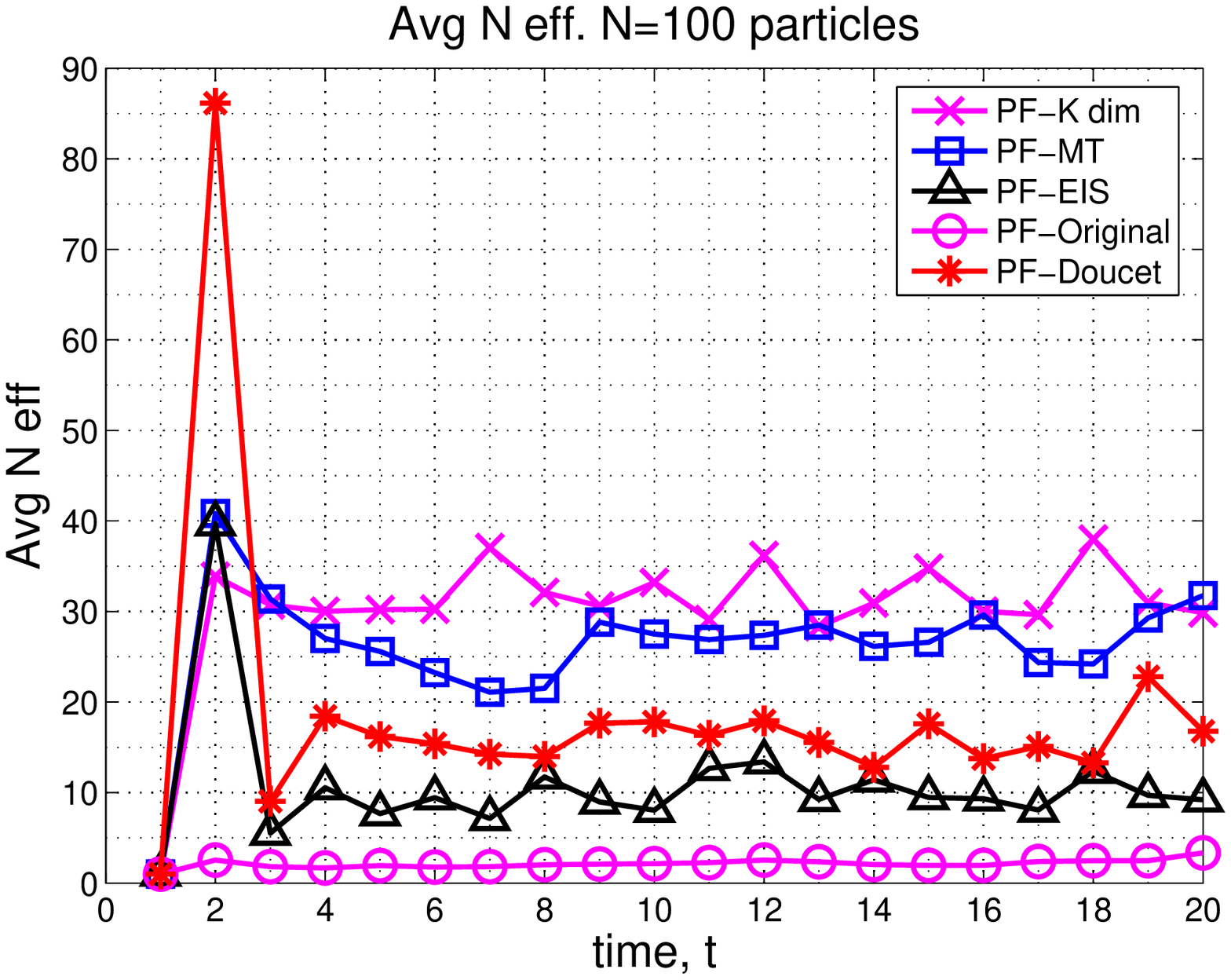, width=4.25cm,height=4cm}
\end{tabular}
}
}
\caption{\small{Effective particle sizes ($N_{eff}$). Because of resampling at each $t$, $N_{eff}$ only measures the effectiveness of the current particles, and not how they influence future posterior estimates. It is high even when most particles cluster around an OL mode which in future turns out to be the wrong one, resulting in larger future RMSEs. PF-Doucet has highest $N_{eff}$, but not lowest RMSE or out-of-track \% (see Fig. \ref{pfeisfigs}).
}}
\label{Nefffigs}
\end{figure*}

Using an anonymous reviewer's suggestion, we also plot the effective particle size, $N_{eff}$, for all the above examples in Fig. \ref{Nefffigs}. $N_{eff}$ is equal to the inverse of the variance of normalized particle weights \cite{tutorial}. Because of resampling at each $t$, $N_{eff}$ only measures the effectiveness of the current particles, and not how they influence the future posterior estimates. $N_{eff}$ will be high even when most particles cluster around an OL mode which in future turns out to be the wrong one, resulting in larger future RMSEs. This is why PF-Doucet, which samples from the Laplace approximation to the ``optimal" importance density (optimal in the sense of minimizing the conditional weights' variance) has the highest $N_{eff}$, but not the smallest RMSE. This issue is most obvious for the squared sensor case.

{\em Time Comparison. } We used the MATLAB profiler to compare the times taken by different PFs for tracking for 20 time steps. 
GSPF-I took 1 second, PF-Original took 2 seconds,  PF-EIS took 60.2 seconds, and PF-Doucet took 111.2 seconds. GSPF-I and PF-Original took significantly lesser time since they do not use gradient descent at all. Note also that the gradient descent algorithm used by us was a very basic and slow implementation using the {\em fminunc} function in MATLAB, thus making PF-EIS or PF-Doucet more slower than they would actually be.
 PF-Doucet takes more time than PF-EIS because (a) it finds the mode on an $M$ dimensional space, while PF-EIS finds mode only on an $M-K$ dimensional space and (b)  $p^*$ is very likely to be multimodal (many times the initial guess particle may not lie in the basin-of-attraction of any mode and so many more descent iterations are required).%


{\em Evaluating  PF-MT and PF-EIS-MT: }
In Fig. \ref{pfmtfigs}, we compare the performance of PF-MT and PF-EIS-MT with other PFs. The model of Fig. \ref{pfmt3_N100fig} was similar to that of Fig. \ref{pfeisnew2N100}, but with  $M=10$. 
We used $X_{t,s} = v_{t,1}$, $X_{t,r,s}=$ empty and $X_{t,r,r} = v_{t,2:10}$, i.e. this was a  PF-MT with $\tilde{X}_{t,s} = v_{t,1}$ and $\tilde{X}_{t,r} = v_{t,2:10}$. As can be seen from the figure, PF-MT outperforms all other algorithms. It outperforms PF-EIS because it importance samples  only on a $K=1$ dim space, but performs MT on the other 9 dimensions (which have a narrow enough conditional posterior) and so its effective particle size is much higher (see Fig. \ref{Neff4}). This is particularly important when the available $N$ is small. 
PF-MT outperforms PF-Doucet primarily because of the EIS step (approximated by MT). It is much better than PF-Original again because of better effective particle size (result of using EIS instead of IS from STP). Finally, it is significantly better than PF-K-dim because  PF-K-dim performs dimension reduction on 9 states (all of which are nonstationary) which results in very large error, while PF-MT tracks the posterior mode on all these dimensions. Note that because of resampling, $N_{eff}$ may also be very high when a PF is completely out-of-track (all particles have very low but roughly equal weights). This is true for PF-K-dim (Fig. \ref{Neff4}).
%

In Fig. \ref{pfmtrobust}, we evaluate robustness to modeling error in sensor failure probability. The tracker assumed failure probability $\alpha_1^{(1)} = 0.2$. The observations were simulated using $\alpha_1^{(1)} = 0.95$. This simulates the situation where a sensor begins to fail much more often  due to some sudden damage to it. For this problem, $M=5$. We used $X_{t,s}=v_{t,1}$, $X_{t,r,s}=v_{t,2}$ and $X_{t,r,r}=v_{t,3:10}$ i.e. we implemented PF-EIS-MT. PF-EIS-MT has the best performance when $N=50$ (available number of particles is small) while PF-EIS has the best performance when a larger $N$, $N=100$ is used (not shown). 

Note that $M=5$ or 10 is a large enough dimensional state space if reasonable accuracy is desired with as low as $N=50$ or 100 particles. In practical scenarios (which are difficult to run multiple Monte Carlo runs of) such as contour tracking \cite{pami07,cdc06journal} or tracking temperature in a wide area with large number of sensors, the state dimension can be as large as 100 or 200 while one cannot use enough particles to importance sample on all dimensions. The IS-MT approximation will be really useful for such types of problems.

\section{Discussion and Future Work}
\label{discussion}

We have studied efficient importance sampling techniques for PF when the observation likelihood (OL) is frequently multimodal or heavy-tailed and the state transition pdf (STP) is broad and/or multimodal. The proposed PF-EIS algorithm generalizes Doucet's idea of sampling from a Gaussian approximation to the optimal importance density, $p^*$, when $p^*$ is unimodal, to the case of multimodal $p^*$.

{\em Sufficient conditions to ensure unimodality of  $p^*$ conditioned on the ``multimodal states", $X_{t,s}$, are derived in Theorem \ref{unimodthm}.  Theorem \ref{unimodthm} can be extended to test for unimodality of any posterior.} Specifically, it can also be extended to problems involving static posterior importance sampling. In its current form, it is very expensive to verify the conditions of  Theorem \ref{unimodthm}. But, based on it, multiple heuristics to choose $X_{t,s}$ to ensure that $p^*$ conditioned on $X_{t,s}$ is most likely to be unimodal have been proposed. An unsolved research issue is to either find efficient numerical techniques to verify the conditions of Theorem \ref{unimodthm} on-the-fly or to find ways to modify the result so that the selection can be done a-priori.
%
%

We have shown through extensive simulations that PF-EIS outperforms PF-Doucet (PF-EIS with $K=0$) whenever $p^*$ is frequently multimodal. But, in other cases, PF-Doucet has lower error. An efficient algorithm (in terms of the required $N$) would be to choose the dimension and direction of $X_{t,s}$ on-the-fly using Heuristic \ref{onfly}.

Increasing $N$ for any PF increases its computational cost. Once $X_{t,s}$ is large enough to satisfy unimodality w.h.p., the $N$ required for a given error increases as dimension of $X_{t,s}$ is increased further (for e.g., PF-Original had much higher RMSE than PF-EIS for given $N$). But, computational cost per particle always reduces as dimension of $X_{t,s}$ is increased (for e.g. PF-Original took much lesser time than PF-EIS which took lesser time than PF-Doucet). {\em For a given tracking performance, if one had to choose $X_{t,s}$ to ensure minimal computational complexity, then the optimal choice will be a higher dimensional $X_{t,s}$ than what is required to just satisfy unimodality.} Finding a systematic way to do this is an open problem. On the other hand, if the goal was to find a PF with minimal storage complexity or to find a PF that uses the smallest number of parallel hardware units (in case of a parallel implementation), the complexity is proportional to $N$. In this case, PF-EIS (or PF-EIS-MT) with smallest possible ``multimodal state" dimension would be the best technique.

As state space dimension increases, the effective particle size reduces (variance of weights increases), thus making any regular PF impractical for large dimensional tracking problems. The posterior Mode Tracking (MT) approximation to importance sampling (IS) for the states whose conditional posterior is narrow enough, is one way to tackle this issue. The IS-MT approximation introduces some error in the estimation of these states, but at the same time, it also reduces the sampling dimension by a large amount, thus improving effective particle size. For carefully chosen IS-MT directions, the net effect is smaller total error, especially when the available $N$ is small.
%
%
An open issue is to find rigorous techniques to select the IS-MT directions to ensure maximum reduction in error. A related issue is to study the stability of PF-MT or PF-EIS-MT, i.e. to show that the increase in PF error due to the IS-MT approximation at a certain time $t_0$ goes to zero with $t$ fast enough and thus the net error due to IS-MT at all times is bounded. A related work is \cite{chopin} which analyzes the RB-PF. An interesting open question is if Compressed Sensing \cite{candes} can be used to select the IS-MT directions and when.

\vspace{-0.05in}
\section*{Acknowledgements}
\vspace{-0.05in}
The author would like to thank the anonymous reviewers and the associate editor for their excellent comments and suggestions. The author would also like to thank Aditya Ramamoorthy for useful comments on Theorem \ref{unimodthm}.
\vspace{-0.075in}

\bibliographystyle{IEEEbib}
\bibliography{visual-tracking-bib,levelsets-bib,occlusion-bib,tip,books}


\end{document}